\documentclass[twocolumn]{emulateapj}
\usepackage{apjfonts}
\usepackage{amsmath}
\usepackage{amssymb}
\usepackage{amstext}
\usepackage{graphicx}
\usepackage{epsfig}
\usepackage{color}

\begin{document}

\shorttitle{Optical models of ULXs}
\shortauthors{Madhusudhan, et al.}
\title{Models for the Observable System Parameters of Ultraluminous 
X-ray Sources}

\author{N. Madhusudhan\altaffilmark{1}, S. Rappaport\altaffilmark{1}, 
Ph. Podsiadlowski\altaffilmark{2} \& L. Nelson\altaffilmark{3} }

\altaffiltext{1}{Department of Physics and Kavli Institute for 
Astrophysics and Space Research, MIT, Cambridge, MA 02139; {\tt 
nmadhu@mit.edu}}
\altaffiltext{2}{Department of Astrophysics, Oxford University, 
Oxford OX1 3RH, UK}
\altaffiltext{3}{Physics Department, Bishop's University, 
Sherbrooke, QC Canada J1M 0C8}

\begin{abstract}

We investigate the evolution of the properties of model populations 
of ultraluminous X-ray sources (ULXs), consisting of a black-hole accretor 
in a binary with a donor star. We have computed models corresponding 
to three different populations of black-hole binaries, motivated by 
our previous studies. Two of the models invoke stellar-mass 
($\sim10\,M_\odot$) black-hole binaries, generated with a binary 
population synthesis code, and the third model utilizes intermediate-mass 
($\sim1000\,M_\odot$) black-hole accretors (IMBHs). For each of the 
three populations, we computed 30,000 binary evolution sequences using
a full Henyey stellar evolution code. A scheme for calculating the
optical flux from ULXs by including the reprocessed X-ray irradiation
by, and the intrinsic viscous energy generation in, the accretion
disk, as well as the optical flux from the donor
star, is discussed. We present color-magnitude diagrams (CMDs) as
``probability images'' for the binaries as well as for the donor stars
alone. ``Probability images'' in the plane of orbital period and X-ray
luminosity are also computed.  We show how a population of
luminous X-ray sources in a cluster of stars evolves with time.  
The most probable ULX system parameters
correspond to high-mass donors (of initial mass $\gtrsim 25~M_\odot$)
with effective O through late B spectral types and equivalent luminosity
classes of IV or V. We also find the most probable orbital periods of
these systems to lie between 1-10 days.  Estimates of the numbers of
ULXs in a typical galaxy as a function of X-ray luminosity are also
presented.  From these studies we conclude that if the stellar-mass
black-hole binaries are allowed to have super-Eddington limited X-ray
luminosities: (i) the value of the binding energy parameter for the
stellar envelope of the progenitor to the black-hole accretor must be
in the range of $0.01 \lesssim \lambda \lesssim 0.03$ in order not to
overproduce the ULXs, and (ii) the stellar-mass black-hole models still have 
a moderately difficult time explaining the observed ULX positions in the 
CMD.  Other possible explanations for the apparent overproduction
of very luminous X-ray sources in the case of stellar-mass black-hole accretors 
are discussed. Our model CMDs are compared with six ULX counterparts that have been
discussed in the literature. The observed systems seem more closely
related to model systems with very high-mass donors
in binaries with IMBH accretors.  We find that a significant 
contribution to the optical flux from the IMBH systems comes from {\em intrinsic} 
accretion disk radiation whose source is viscous dissipation of gravitational potential
energy.  In effect, the IMBH systems, when operating at their maximum luminosities 
($10^{41}-10^{42}$ ergs s$^{-1}$), are {\em milli-AGN}.  With regard to the IMBH 
scenario, while attractive from the aspect of binary evolution models, it leaves 
open the larger question of how the IMBHs form, and how they capture massive 
stellar companions into just the correct orbits.

\end{abstract}


\section{Introduction}
\label{sec:intro}

An ultraluminous X-ray source (ULX) is an off-nucleus point-like 
source whose X-ray luminosity, $L_x$, exceeds $2 \times 10^{39}$ ergs 
s$^{-1}$. An obvious candidate for such a source is an X-ray binary 
where a compact object accretes from a donor star. Under normal 
circumstances, the accretion rate is constrained by the Eddington 
limit of the accretor. The Eddington limit for a neutron star is 
$\sim$ $10^{38}$ ergs s$^{-1}$, and that for a $10\,M_\odot$ black 
hole (BH) is $\sim$ $10^{39}$ ergs s$^{-1}$. Therefore, if the 
accretors are assumed to be stellar mass BHs, the observed 
luminosities of ULXs exceed the Eddington limit by factors of up to $\sim$100. 
The question as to how the Eddington limit could be violated 
remains a subject of considerable debate. On the other hand, if we 
assume the accretor to be an intermediate-mass black hole of $\sim$ 
$10^2 - 10^4\,M_\odot$, the ULX luminosities can all be accounted for 
without violating the Eddington limit. If the accreting stars are 
indeed IMBHs, the question remains as to how these objects are formed 
and how they acquire a massive stellar companion.

Several scenarios have been put forth to explain the apparent
super-Eddington luminosities, assuming stellar-mass BHs. King et
al.~(2001) suggested that geometrical beaming in the direction of the
observer, due to a thick accretion disk (Jaroszy\'nski, Abramowicz, \& 
Paczy\'nski 1980), could lead to both a high luminosity and an even
higher inferred luminosity.  K\"ording et al.\,(2002) make a similar argument 
with relativistic beaming, due to jets, as the explanation. In both of
these scenarios, the emission is considered to be anisotropic.  These
somewhat contradict the observations of ionization nebulae around some
ULXs (Pakull \& Mirioni 2003) which tend to indicate the full implied
isotropic luminosity. Another scenario proposed by Begelman
(2002; 2006) suggests that photon bubble instabilities in the
accretion disk can lead to an isotropic luminosity exceeding the
Eddington limit by a factor of $\sim$10. Socrates \& Davis (2006)
invoke a hot, optically thin corona in conjunction with a
geometrically thin, but optically thick, accretion disk to explain the
observed ULX luminosities. Yet other scenarios invoke ``slim'' 
accretion disks to produce approximately isotropic luminosities that
may exceed the Eddington limit by up to factors of $\sim$10 (Abramowicz 
et al.~1988; Ebisawa et
al.\,2003). Although these scenarios have not been shown to work
conclusively, they could, in principle, account for ULX luminosities
up to $\sim$$10^{40}$ ergs s$^{-1}$.  In addition to these specific
suggestions for bypassing the Eddington limit, the continuity and
simplicity of the luminosity function of luminous X-ray sources from
$10^{36}-10^{40}$ ergs s$^{-1}$ (a simple power law; Grimm et
al.~2003) has led some to conclude that these represent a single class
of systems, i.e., neutron stars and stellar-mass black holes\footnote{
In this paper, we refer to a stellar-mass black holes as a ``low-mass
black hole'' or ``LMBH'', rather than ``SMBH'' which might be
confused with ``supermassive black hole''.}
accreting from a normal donor star.  However, we note that essentially
none of these ideas involving stellar-mass BHs would plausibly be able to
explain the ULXs at the higher end of the observed luminosities,
namely those with $L_x \gtrsim 10^{40}$ ergs s$^{-1}$.

By contrast, the IMBH scenario, as suggested first by Colbert \&
Mushotzky (1999), accounts for the luminosities of all the ULXs
because the Eddington limit for an IMBH in the mass range $\sim$ $10^2
- 10^4\,M_\odot$ varies between $10^{40} - 10^{42}$ ergs
s$^{-1}$. Other evidence that may suggest the presence of an IMBH
includes observations of mHz QPOs (Strohmayer \& Mushotzky 2003) and
ionization nebulae (Pakull \& Mirioni 2003), and inferences of cool
inner accretion disk temperatures (Miller et al.~2003; Miller et
al.~2004; Cropper et al.~2004). Despite the several pieces of evidence
supporting IMBH binaries as the model for ULXs, one of the main
problems with this scenario lies in the unknown formation mechanism
for such binaries. A few scenarios for IMBH formation have been
proposed in the literature (see, for example, Portegies Zwart et
al.~2004; Tutukov \& Fedorova 2005) but a definitive scenario remains
elusive. In particular, the evolution of supermassive stars with 
$M \gg 100~M_\odot$ is not well understood.

In three of our earlier papers, we studied the evolution of 
populations of binary systems pertaining to both models for ULXs, 
namely: (i) super-Eddington accretion onto LMBHs (Podsiadlowski et 
al.\,2003, Rappaport et al.\,2005); and, (ii) sub-Eddington accretion 
onto IMBHs (Madhusudhan et al.\,2006). In those studies, the main 
objective was to investigate the evolution of the X-ray luminosities 
of the systems with time, and to estimate the formation efficiencies 
for each scenario. In Podsiadlowski et al.~(2003) and Rappaport et 
al.~(2005), we considered LMBHs in binary systems with donor stars in 
the mass range $2-17\,M_\odot$. Some of the binary populations were 
generated using the binary population synthesis (BPS) code developed 
in Podsiadlowski et al.~(2003). There, we showed that by allowing 
for a violation of the Eddington limit by a factor of $\sim$$10-30$, 
such systems would be able to explain most of the observed ULXs, 
except for the very most luminous ones. In Madhusudhan et al.~(2006), 
we considered donor stars in binary systems with IMBHs with a 
representative mass of $1000\,M_\odot$. There, we showed that in 
order to have a plausible formation efficiency for ULXs with $L_{x} 
\gtrsim 10^{40}$ ergs s$^{-1}$ the donor stars should be massive 
($\gtrsim 8~M_\odot $)  and the initial orbital separations should be 
close ($\lesssim 6-40$ times the radius of the donor star when on the 
main sequence).

\begin{deluxetable}{l c c c}
\tablewidth{0pt}
\tablecaption{Magnitudes and Colors of ULX counterparts \newline (adapted from Copperwheat et al. 2007)}
\label{tab:obsdata}
\tablehead{\colhead{System} & \colhead{$M_{\rm B}$} & \colhead{$M_{\rm V}$} & \colhead{$B-V$}}

\startdata
NGC 4559 X-7 (C1) & $-7.22 \pm 0.19$&$-7.03 \pm 0.16$&$-0.19 \pm 0.25$ \\
M81 X-6 & $-4.28 \pm 0.04$ & $-4.18 \pm 0.03$ & $-0.10 \pm 0.05$ \\
M101 ULX-1        & $-6.19 \pm 0.15$ & $-5.92 \pm 0.12$ & $-0.27 \pm 0.19$ \\
NGC 5408 ULX      & $-6.40 \pm 0.20$ & $-6.40 \pm 0.20$ & $-0.00 \pm 0.28$ \\
Holmberg II ULX   & $-6.03 \pm 0.19$ & $-5.78 \pm 0.11$ & $-0.25 \pm 0.22$ \\
NGC 1313 X-2 (C1) & $-4.70 \pm 0.18$ & $-4.50 \pm 0.18$ & $-0.20 \pm 0.25$ \\
\enddata

\end{deluxetable}

A number of fairly secure optical identifications of ULX counterparts 
have now been reported in the literature. Liu et al.\,(2002) 
identified a unique counterpart to the ULX NGC 3031 X11 (in M81) and 
found it to be consistent with an O8 main sequence (MS) donor star. 
Kaaret et al.\,(2004) reported an optical counterpart to the ULX in 
Holmberg II to be consistent with a donor star with spectral type 
between O4 V and B3 Ib. Kuntz et al.~(2005) have studied the optical 
counterpart of M101 ULX-1 and found the colors to be consistent with 
those for a mid-B supergiant. Soria et al.\,(2005) studied candidate 
counterparts to the ULX in NGC 4559 and suggested the donor to be 
either a blue or a red supergiant of high mass ($\sim$$10 - 20 
\,M_\odot$). Mucciarelli et al.~(2005; 2007) have found candidate 
optical counterparts to the ULX NGC 1313 X-2 that are either B0-O9 
main-sequence stars or G supergiants. 

The best candidate optical counterparts to the ULXs discussed 
above have been conveniently selected and 
summarized by Copperwheat et al.\,(2007). In particular, they 
tabulate the photometric values for the various systems, calculating 
reddening corrections and absolute magnitudes, wherever necessary. As 
can be seen from the above list, the observations seem to indicate 
that the spectral classes of the most promising candidates generally range 
between O and B spectral type, and the luminosity classes are either V or Ib. 
In the present study, we consider optical counterparts of six ULX 
systems for which photometric data in the B and V bands are available 
in the literature. The photometric data for these systems are given 
in Table 1 (as adapted from Copperwheat et al.\,2007). 

Theoretical studies of ULXs in the optical regime have heretofore 
concentrated on constraining  the nature of the donor star, the mass 
of the accretor, and the orbital period.  All the models follow the 
standard paradigm of ULXs being X-ray binaries with active accretion 
through Roche-lobe overflow from a donor star onto a BH accretor. 
Rappaport et al.~(2005) discussed preliminary theoretical models 
involving disk irradiation in ULXs.
They presented sample evolution tracks, on a color-magnitude diagram, 
for four ULX models, and discussed the optical appearance of ULX BH 
binaries. Pooley \& Rappaport (2005) suggested detection of X-ray 
and/or optical eclipses as a means to constrain the mass of the 
accretor. In recent studies, Copperwheat et al.\,(2005; 2007) have 
also constructed irradiation models to describe the optical emission 
from ULXs, and used the models to constrain the properties of several 
systems observed in the optical. The parameters being considered were 
the mass, radius and age of the donor, and the BH mass in some cases. 
 From the fits to the various observations, they find the counterparts 
to be consistent with being main-sequence stars or evolved 
giants/supergiants with spectral types O, B, or A.

In the present paper, we report a detailed population study spanning 
a large region of parameter space of ULX properties in an effort to 
better explore and help constrain ULX models. In particular, we investigate 
models of the optical properties of ULXs. We choose three sets of 
representative populations from our previous studies, and explore 
their optical properties and other observables. For the models with 
LMBHs, we choose two sets of populations obtained from the BPS code 
corresponding to two different prescriptions for the critical mass ratio in
the primordial progenitor binary that determines whether a common
envelope phase can occur.  These two models yield very different distributions
of secondary (i.e., ULX donor) star masses.
For models with IMBHs, we choose model C from Madhusudhan et 
al.~(2006). This model yielded the highest formation efficiency for 
ULXs in the IMBH scenario. For, each of the three populations, we 
follow 30,000 binary evolution calculations.  The models are summarized 
in Table 2.

The X-ray luminosities are calculated using a standard formulation of 
Roche-lobe mass transfer in X-ray binaries. The optical flux from the 
system is determined as the sum-total of the optical flux from the 
donor star, X-ray irradiation of the accretion disk, and intrinsic energy 
generation in the disk. We present detailed 
color-magnitude diagrams (CMDs) for all the models. These include 
CMDs for the binary system, i.e., the sum of the optical flux of the 
donor and that due to radiation from the disk, as well as CMDs for 
the donor star alone. We also study in detail the evolution of the 
X-ray luminosity with the age of the system, $t_{\rm ev}$, and the 
evolution of the X-ray luminosity with orbital period, $P_{\rm orb}$.

\section{Methods}
\label{sec:methods}

\subsection{Binary Population Synthesis}

\begin{figure}[h]
\begin{center}
\includegraphics[width=0.48\textwidth]{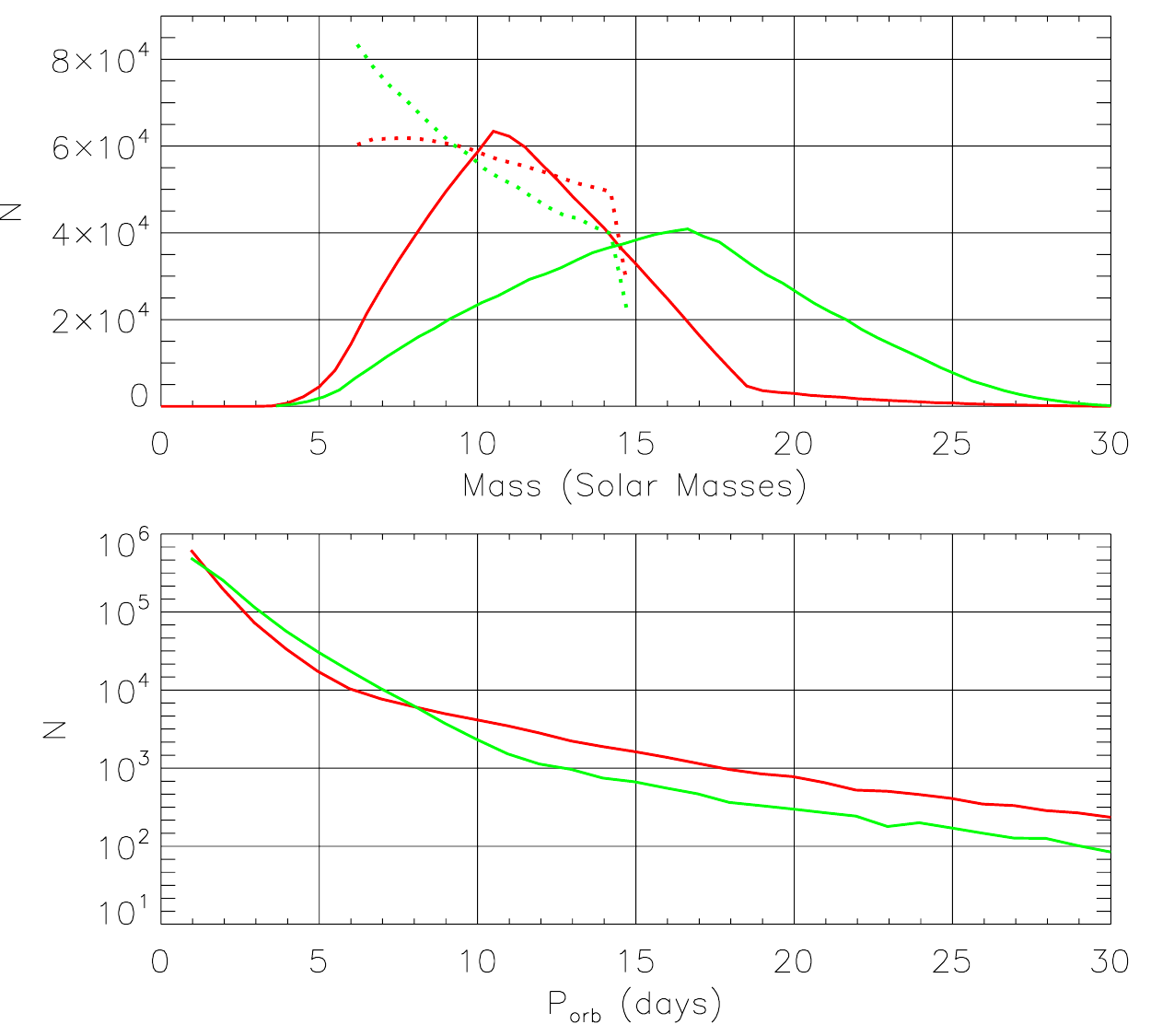}
\caption{Distributions of initial donor masses, black hole masses and 
initial orbital periods of the LMBH systems. The red curves 
correspond to Model La and the green curves correspond to Model Lb. 
In the top panel, the solid curves represent donor masses and the 
dashed curves represent BH masses. For Model Ic, $M_{\rm BH} = 
1000~M_\odot$ and we choose $M_{\rm don}$ uniformly between $5-50 
~M_\odot$, and the initial orbital separation uniformly between 1-3 
times the separation required for a ZAMS star of the corresponding 
donor mass to fill its Roche lobe.}
\label{fig:models}
\end{center}
\end{figure}

In our previous studies we explored several models of ULX populations 
for both the low- (hereafter ``LMBH'') and intermediate-mass 
(hereafter ``IMBH'') black-hole scenarios (Rappaport et al.\,2005; 
Madhusudhan et al.\,2006). In the present work, we chose two LMBH- 
and one IMBH-binary population that were motivated by these previous 
studies. The parameters for the three models are summarized in Table 
\ref{tab:modelsdata}, and the distributions of the initial system 
masses and orbital periods are shown in Figure \ref{fig:models}. 
Models La and Lb are LMBH populations, whereas Model Ic represents an 
IMBH population.

To generate the LMBH binary populations, we used the binary 
population synthesis (BPS) code developed in Podsiadlowski et 
al.\,(2003). We briefly review here the formulation from Rappaport et 
al.\,(2005). We started with a very large set of massive primordial 
binaries and generated a much smaller subset of these that evolved to 
contain a black hole and a relatively unevolved companion star.  The 
product was a set of ``incipient'' black-hole X-ray binaries with a 
particular distribution of orbital periods, $P_{\rm orb}$, donor 
masses, $M_{\rm don}$, and black-hole masses, $M_{\rm BH}$, for
each of a number of different sets of input assumptions (see, e.g., 
Fig.\ 2 of Podsiadlowski et al.\,2003).  For this part of the 
calculation, we employed various ``prescriptions'', based on single-star 
evolution models for the primary, simple orbital dynamics 
associated with wind mass loss and transfer, assumptions about the 
magnitude of the wind mass loss from the primary as well as from the 
core of the primary after the common envelope, and natal kicks during 
the core collapse and formation of the black hole.

The decision in the binary population synthesis code of whether or
not a common-envelope phase occurs, when the primary first overflows
its Roche lobe and starts to transfer mass onto the secondary, is based 
on the evolutionary state of the primary when mass transfer commences.
We first define $R_{\rm TAMS}$ and $R_{\rm HG}$ as the radii of the 
primary when it reaches the terminal main sequence and the end of 
the Hertzsprung gap, respectively.  Given the initial orbital separation
and the mass ratio of the primordial binary, $q \equiv M_{\rm prim}/M_{\rm sec}$,
we can compute the Roche-lobe radius, $R_L$, of the primary.  In order
to form a relatively close black-hole binary of the type we are considering
in this work (i.e., $P_{\rm orb} \lesssim 20$ days), we require that the initial 
mass-transfer rate from the primary to the secondary be so large that it leads 
to a common-envelope and spiral-in phase. This depends on the initial mass 
ratio and evolutionary state of the secondary and is somewhat uncertain (see 
Pols 1994; Wellstein, Langer \& Braun 2001).
For one of our two models involving stellar-mass black
hole accretors, we utilize the follow critical mass ratios, $q_{\rm crit}$,
depending on the relation among $R_L$, $R_{\rm TAMS}$, and $R_{HG}$:
\begin{eqnarray}
q_{\rm crit} & = & 2~~~~~~~~{\rm for}~~R_{\rm TAMS} < R_L < R_{HG} \\
q_{\rm crit} & = & 1.2~~~~~{\rm for}~~R_L > R_{HG}
\end{eqnarray}
For the second of our two models involving stellar-mass black-hole
accretors we utilize an interpolated version of the above prescription:
\begin{equation}
q_{\rm crit} = 2.0 - 0.8 \left[ \frac{\log R_L- \log R_{\rm TAMS}}{\log R_{\rm HG}- \log R_{\rm TAMS}} \right] 
\end{equation}
for $R_{\rm TAMS} < R_L < R_{\rm HG}$.  This somewhat ad hoc prescription 
has the net effect of producing a much greater fraction of incipient ULX sources 
with high-mass donor stars.

\begin{deluxetable}{c c c c c c}
\tablewidth{0pt}
\tablecaption{ULX Population Models
\label{tab:modelsdata}}
\tablehead{
\colhead{Model\tablenotemark{a}} & \colhead{$M_{\rm BH}$\tablenotemark{b}} &
\colhead{$M_{\rm don}$\tablenotemark{c}} & \colhead{$P_{\rm orb,i}$\tablenotemark{d}} & \colhead{$\lambda$\tablenotemark{e}} & \colhead{$q_{\rm crit}$\tablenotemark{f}}}
\startdata
La ...............&   $6-15$ &    $5-18$ & $1-18$ & $0.1$ & 1.2 or 2.0 \\
Lb ...............&   $6-15$ &    $5-27$ & $1-12$ & $0.1$ & $1.2 \rightarrow 2.0$ \\
Ic ...............&   $1000$ &    $5-50$ & $1-10$ & $ - $ & $ - $  \\ 
\enddata
\tablenotetext{a}{The distributions of masses and periods for the different models are shown in Figure \ref{fig:models}}
\tablenotetext{b}{Approximate range of black hole masses, in units of $M_\odot$} 
\tablenotetext{c}{Approximate range of donor masses, in units of $M_\odot$}
\tablenotetext{d}{Approximate range of initial orbital periods, in units of days}
\tablenotetext{e}{Dimensionless inverse binding energy of the envelope of the black-hole progenitor star}
\tablenotetext{f}{For details of the prescription see eqns.~(1) $-$ (3) in the text.}
\end{deluxetable}

Simple energetic arguments were used to yield the final-to-initial 
orbital separation during the common-envelope phase wherein the 
envelope of the primary is ejected.  Here we utilized a parameter 
$\lambda$, which is the inverse of the binding energy of the primary 
envelope at the onset of the common-envelope phase in units of 
$GM_{\rm prim} M_e/R_{\rm prim}$, where $M_{\rm prim}$, $M_e$ and 
$R_{\rm prim}$ are the total mass, 
envelope mass, and radius of the primary, respectively.  This 
parameter strongly affects the final orbital separation after the 
common-envelope phase, where smaller values of $\lambda$ correspond to
more tightly bound envelopes, and hence more compact 
post-common-envelope orbits.

With the above parameterization for the ejection of the common 
envelope we find the following expression for initial-final orbital 
separation:
\begin{equation}
\left(\frac{a_f}{a_i}\right)_{\rm CE} \simeq \frac{M_c M_{\rm sec}}{M_{\rm prim}}
\left(M_{\rm sec}+\frac{2M_e}{\mathcal{E}_{\rm CE}\lambda r_{\rm L}}\right)^{-1}~,
\label{bps1}
\end{equation}
(e.g., Webbink 1985; Dewi \& Tauris 2000; Pfahl et al. 2003), where 
the subscripts ``prim'', ``c'', and ``e'' stand for the progenitor of 
the black hole, its core, and its envelope, respectively, and ``sec'' 
is for the secondary, i.e., the progenitor of the ``donor star'' in the black-hole binary. 
The quantity $r_{L}$ is the Roche-lobe radius of the black-hole 
progenitor in units of $a_i$, $\mathcal{E}_{\rm CE}$ is the fraction of the 
gravitational binding energy between the secondary and the core of 
the black-hole progenitor that is used to eject the common envelope, 
and $\lambda$ is defined above.  For typically adopted parameter 
values, $\lambda \sim 0.01-1$ (e.g., Dewi \& Tauris 2000; Podsiadlowski et al.\ 2003), 
$\mathcal{E}_{\rm CE} \simeq 1$, and $r_L \simeq 0.45-0.6$ (for an assumed mass ratio 
between the black-hole progenitor and the companion in the range of 
$\sim 2:1 \rightarrow 15:1$), the second term within the parentheses 
in eq.\,(\ref{bps1}) dominates over the first.  In this case, we find 
the following simplified expression for $a_f/a_i$:
\begin{equation}
\left(\frac{a_f}{a_i}\right)_{\rm CE} \simeq \frac{r_{L}}{2}
\left(\frac{M_c}{M_{\rm prim}M_e}\right) M_{\rm sec} \lambda \simeq 0.005 
\left(\frac{M_{\rm sec}}{M_\odot}\right)
\lambda~,
\label{bps2}
\end{equation}
where the leading factor is $r_{L}/2 \simeq 1/4$, while the factor in 
parentheses involving the black-hole progenitor is $\sim 0.020 \pm 
0.002~M_\odot^{-1}$ for virtually all of the progenitors we consider. 
This explains why the large majority of the incipient black-hole 
binaries (with low- to intermediate-mass donors) 
found by Podsiadlowski et al.~(2003) resulted from an 
initially very wide orbit ($P_{\rm orb} \sim ~ $years -- when the 
primary attains radii of $\sim$$1000-2300~R_\odot$) preceding the 
common-envelope phase in order to avoid a merger between the 
secondary and the core of the primary.

In the present work, for Models La and Lb (see Table \ref{tab:modelsdata} and 
Fig.\,\ref{fig:models}), we chose two populations that were obtained using the 
BPS code described above. These two LMBH populations both utilize 
an inverse envelope binding energy parameter $\lambda = 0.1$, and 
two different prescriptions for the critical mass ratio required to produce
a common envelope phase (see eqns.~[1]--[3]).  These result in quite 
different production rates for incipient ULXs with high-mass donor stars
(i.e., $\gtrsim 15~M_\odot$).

For the IMBH case, Model Ic (see Table \ref{tab:modelsdata}), we 
chose the same population as Model C in Madhusudhan et al.\,(2006). 
Due to our lack of a firm understanding of how IMBHs form in star 
clusters, and how they capture companion stars, we adopted a very 
simple prescription for the population of incipient IMBH binaries 
(but, see Blecha et al.~2006). For each binary, we chose the initial
donor mass to lie uniformly in the range of $5-50$ $M_\odot$ using
Monte Carlo methods. And, we chose the initial orbital separation 
uniformly over the range of $1-3$ times the separation required for a 
ZAMS star of the corresponding mass to fill its Roche lobe. This 
translates to a range of initial orbital separations between $40-200$ 
$R_\odot$, and $P_{\rm orb}$ in the range of $\sim$$1-10$ days. 
We found in our previous work that this population is the 
most favorable one we explored for producing ULXs in the IMBH 
scenario, i.e., the donor stars should be massive, i.e., $\gtrsim 8 
~M_\odot$ and the initial orbital separations, after circularization, 
should be close, $\sim$ $6-40$ times the radius of the donor star when 
on the ZAMS. This range of close orbital separations represents the case 
of direct tidal capture and circularization of a single field star 
by the IMBH (see, e.g., Hopman et al.\ 2004; Pfahl 2005).

For, each of the three modeled ULX populations, we carried out 30,000 
binary evolution calculations. All models were run using 60 nodes of the {\tt 
elix3} Beowulf cluster located at the University of Sherbrooke, 
Quebec.  The run time for each of the three models was $\sim$40 hours.

\subsection{Stellar Evolution}

The stellar evolution of the donor stars, including mass loss, was 
followed with
{\tt EZ} which is a stripped down, rewritten version of a subset of 
the stellar evolution code developed by P. P. Eggleton (Paxton 2004). 
The physics of the program is unchanged from Eggleton's (essentially 
as described in Pols et al.\,1995), but the structure of the code has 
been modified to facilitate experiments involving programmed control 
of parameters.  There are zero-age main-sequence (ZAMS) starting 
models for a variety of metallicities (from $Z=10^{-4}$ to $Z=0.03$) 
and masses (from 0.1 to 100 $M_\odot$), with arbitrary
starting masses created by interpolation.  A user-provided procedure is called
between steps of the evolution to inspect the current state, to make changes in
parameters, and to decide when and what to record to log files. The 
source code and data for {\tt EZ} can be downloaded from the web at 
$<$http://theory.kitp.ucsb.edu/$\sim$paxton$>$.  For all models in 
this particular study, the number of stellar mesh-points was fixed at 
200 in the interest of minimizing computation time.

\subsection{Binary Evolution Calculations}

The binary evolution was governed by a sequence of calculations
involving the mass-transfer rate, the corresponding change in orbital
separation, and the subsequent monitoring of mass loss. The mass
transfer considered was due solely to Roche-lobe overflow. When the
donor star fills its Roche lobe, the excess matter flows through the
inner Lagrange point onto the accretor.  Assuming spherical geometry,
the mass-transfer rate was calculated using $\dot M \simeq 2\pi R H
\rho v$, where $R$ is the radius of the donor, $H$ is the density
scale height of the donor atmosphere, $\rho$ is the density of the
atmosphere at the Roche lobe, and $v$ is the thermal velocity at the
photosphere. Under the approximation of an isothermal,
constant-gravity atmosphere, the scale height is given by $H \simeq
kT/\mu g$, where $\mu$ is the mean molecular weight, and the 
density profile of the atmosphere is exponential.
The Roche-lobe radius of the donor was taken to be $R_{\rm RL} =
0.49\,a\,q^{2/3}[0.6\,q^{2/3} +\ln\,(1+q^{1/3})]^{-1}$ (Eggleton
1983), where, $a$ is the separation of the binary, $q = M_{\rm
don}/M_{\rm BH}$, $M_{\rm don}$ is the mass of the  
donor, and $M_{\rm BH}$ is the mass of the black hole.  
This procedure, while making use
of some approximations, is self-adjusting to yield the correct
mass-transfer rates.

The Eddington limited mass-transfer rate onto the accretor is given 
by $\dot M_{\rm Edd} = 4\pi G M_{acc}/\eta \kappa c$, where $G$ is the 
gravitational constant, $\kappa$ is the radiative opacity, and $\eta$ 
is the efficiency of the black-hole accretor in converting rest mass 
to radiant energy. The opacity is assumed to be predominantly due to 
electron scattering and is given by $\kappa = 0.2 (1+X)$ cm$^2$ 
g$^{-1}$, where, $X$ is the hydrogen mass fraction. For a spinning 
black-hole, we take the accretion efficiency to be given by $\eta = 
1 - \sqrt{1 - (M/3 M_0)}$, where $M_0$ is the initial mass of the 
black-hole (Bardeen 1970). This assumes an initially non-spinning black-hole.

If we consider all the matter leaving the donor to be retained by the 
accretor, independent of the Eddington limit, then the resulting 
luminosity is referred to as the ``potential luminosity'' and is 
given by, $L_{\rm pot}=\eta \dot M c^2$. 
The actual isotropic X-ray luminosity, on the other hand, is given by $L_x = 
\beta \eta \dot M c^2$, where $\beta$ restricts the mass-transfer 
rate to the Eddington limit. For $\dot M < \dot M_{\rm Edd}$, $\beta 
= 1$, otherwise $\beta = \dot M_{\rm Edd}/\dot M$. When $\dot M > 
\dot M_{\rm Edd}$, it is assumed that all the matter passes through 
the accretion disk until it reaches the inner edge of the disk, at 
which point the excess matter (above the Eddington limit) is ejected 
out of the system in the form of a jet. The mass and orbital angular 
momentum lost from the system in such ejection is incorporated when 
calculating the orbital separation of the system. We emphasize that we 
use the Eddington limited mass-transfer rate while calculating the 
orbital parameters during the evolution of the binary. However, along 
with the various orbital parameters, we also record $L_{\rm pot}$ at 
each step. And, in all the results presented in this paper, we 
typically refer to $L_{\rm pot}$ instead of the X-ray luminosity, 
$L_x$, of the system in order to examine the potentialities of 
violating the Eddington limit.

\subsection{Disk Irradiation}

The optical flux from a ULX is comprised of contributions from the 
donor star and from the reprocessing of X-ray photons by the 
accretion disk (as well as intrinsic energy generation within the disk). 
The irradiation of the accretion disk depends upon 
$L_x$, the geometrical properties of the disk, and whether or not the 
central accreting star has a hard surface. For purposes of the 
current calculation, we assume the disk to be geometrically thin. We 
find the effective temperature of the disk to be:
\begin{equation}
T(r) \simeq \left(\frac{L_x}{4 \pi r_{\rm min}^2 \sigma}\right)^{1/4} 
\left[\frac{4}{7} \xi'^2 x^{-10/7} (1-\alpha) + 3 \, x^{-3}\right]^{1/4}.
\label{eqn1}
\end{equation}

Here, $x = r/r_{\rm min}$, where $r$ is the radial distance, $r_{\rm
min}$ is the inner disk radius, taken to be $6GM_{\rm BH}/c^2$, $\alpha$
is the X-ray albedo of the disk, which we take to be 0.7.  The half
thickness of the disk is given by $h(r) = \xi r^{9/7}r_{\rm max}^{-2/7}$,
and $\xi' = \xi (r_{\rm min}/r_{\rm max})^{2/7}$, where $\xi$ is a
constant equal to $h(r_{\rm max})/r_{\rm max}$ and $r_{\rm max}$ is
the outer radius of the disk. We have taken $r_{\rm max}$ to be
$0.7r_La$ where $r_L$ is the dimensionless Roche-lobe radius of the
black-hole accretor.  We somewhat arbitrarily adopt a value for $\xi$ of 0.1, 
corresponding to a full angular thickness of the disk of $\sim 12^\circ$. The 
second term in the square brackets in eq.\,(\ref{eqn1}) comes from the 
Shakura-Sunyaev (1973) solutions for the temperature profile of a thin disk 
around a BH accretor. We have taken the inner edge of the accretion disk to 
be at the innermost stable circular orbit around a non-rotating BH.  We have 
neglected the factor ($1-\sqrt{r_{\rm min}/r}$) since its contribution to the optical 
flux can be assumed to be negligible. The first term in the square brackets in 
eq.\,(\ref{eqn1}) is adapted from the expression for the irradiation 
temperature of an accretion disk around a BH derived in King, Kolb \& 
Szuszkiewicz (1997). This is a modified version of the corresponding 
term in eq.\,(8) of Rappaport et al.\,(2005) where we used an expression 
appropriate for 
radiation emanating from a centrally located, isotropically emitting, 
hard surface. The present formulation takes into account the fact 
that the X-ray radiation comes from the innermost part of the accretion 
disk surface and is emitted with a Lambertian angular distribution. 
Since the irradiated portion of the outer disk lies at a shallow 
angle with respect to the surface of the inner disk, the small value 
of the cosine factor significantly reduces the X-ray irradiation. 

As an important caveat to our disk irradiation calculations, we note that 
the assumption of a thin disk (emitting in a Lambertian manner) likely breaks 
down at accretion rates slightly or greatly exceeding Eddington.  In this case, 
the irradiation of the outer disk might become more {\em or} less efficient.  However, 
such a calculation is considerably beyond the scope of the present work.

The optical flux from the disk is determined by using the temperature 
profile described in eq.\,(\ref{eqn1}), and assumes the local 
spectrum in each annulus of the disk is thermal. The radiant flux 
from each annular ring on the disk surface is determined from 
the specific intensity of the blackbody 
radiation and the temperature at that radius. The total flux due to 
the disk at a particular wavelength is calculated by integrating the 
annular flux over the radial extent of the disk, from $r_{\rm min}$ 
to $r_{\rm max}$. The flux from the donor star is determined from its surface 
temperature, $T_e$, as obtained from the stellar evolution code. For the 
wavelengths under consideration, a thermal blackbody is a reasonable 
approximation in order to determine the flux from the star (with 
$\sim$10\% accuracy). This corresponds to an error in the apparent 
$V$ magnitude of $\sim \pm$ 0.1, which is within the error bars of 
most observations, and negligible compared to the absolute values of 
$V$ and $M_V$. The corresponding error in $B-V$ is $\lesssim$ 0.2.

The total flux of the system consists of contributions from the disk 
and the donor star. The $B$ and $V$ magnitudes were calculated from the 
total fluxes using  $B = -2.5\;\log(F_B/F_{B0})$ and $V = 
-2.5\;\log(F_V/F_{V0})$, where $F_B$, $F_{B0}$, $F_V$ and $F_{V0}$ 
are the fluxes and reference fluxes at $\lambda = 4380$ \AA ~and 
$\lambda = 5450$ \AA, respectively ($F_{B0} = 6.61 \times 10^{-9}$ 
ergs cm$^{-2}$ s$^{-1}$ \AA$^{-1}$ and $F_{V0} = 3.64 \times 10^{-9}$ 
ergs cm$^{-2}$ s$^{-1}$ \AA$^{-1} $).

After we compute $B-V$ for the combined radiation from the donor
star and the accretion disk\footnote{All of our binary systems are 
assumed to be viewed at an average orbital inclination angle of $60^\circ$.}, we 
apply a small correction ($\lesssim
0.2$ magnitudes) as a function of $B-V$ to take into account the
treatment of the radiation as blackbody emission.  These corrections 
were based on the differences between the $B-V$ colors computed from
the simple algorithm given above and the colors of main sequence and 
supergiant stars tabulated by Johnson (1966).

\subsection{Intrinsic Disk Emission}

\begin{figure}[t]
\includegraphics[width=1.0\columnwidth]{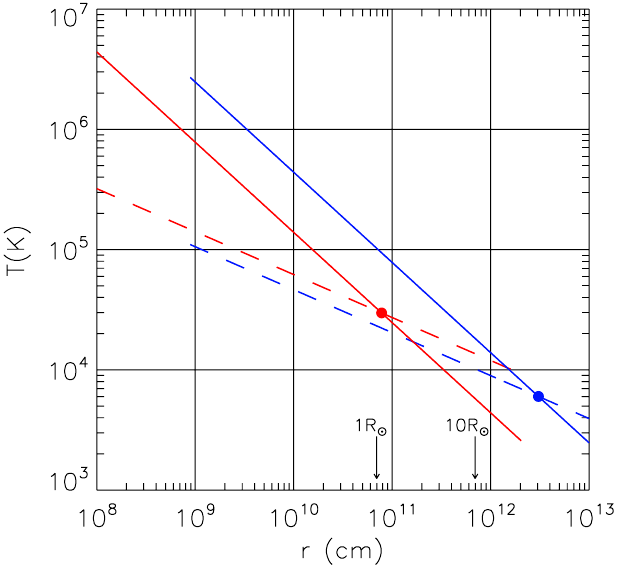}
\caption{Illustrative temperature profiles of the irradiated accretion disk for an 
IMBH model (blue curves) and LMBH model (red curves).  The solid curves are
for the case of intrinsic (viscous) heating of the accretion disk alone, while the dashed
curves are for the irradiation of the disk without any contributions from viscous heating.
These correspond to the 2nd and the 1st terms, respectively, in the square brackets of
eq.~(6). The binary system parameters used for this illustrative example are given in
the text.}
\label{fig:temp}
\end{figure}

We have found from our models (see \S3) that under certain 
circumstances the intrinsic disk emission (due to the viscous release of 
gravitational potential energy) which
is represented by the second term in square brackets in eq.~(6), can 
dominate the optical emission from the disk irradiation (first term in square
brackets).  To illustrate this effect, we show in Fig.\,\ref{fig:temp} the temperature
profiles for an illustrative set of binary parameters: $L_{\rm pot} = 10^{40}$ ergs
s$^{-1}$, $P_{\rm orb} = 30$ days, and $M_{\rm don} = 10~M_\odot$.  For
an IMBH model ($M_{\rm BH} = 1000~M_\odot$) the minimum and maximum
disk radii are $r_{\rm min} = 9 \times 10^8$ cm and $r_{\rm max} = 1.5 \times 
10^{13}$ cm, respectively, while for an LMBH model (with $M_{\rm BH} = 10~M_\odot$), 
these radii are $r_{\rm min} = 9 \times 10^6$ cm, and $r_{\rm max} = 2 \times 
10^{12}$ cm. 

The red and blue curves in Fig.\,\ref{fig:temp} are for the LMBH 
and IMBH models, respectively.  The dashed curves are the temperature 
profiles for the case where irradiation is the only source of energy input to 
the disk; conversely, the solid curves are for the intrinsic (viscous) release
of gravitational energy alone -- without X-ray irradiation.  For the LMBH model, 
we see that the
intrinsic energy release dominates the irradiation contribution to the optical
radiation only for radial distances in the disk of $\lesssim 1~R_\odot$.  Since the 
optical light from the donor star itself comes from a typically much larger area, the
intrinsic disk emission is generally not competitive with that from disk irradiation.  
By contrast the crossing
point for the two temperature profiles corresponding to the IMBH model 
occurs at a radial distance of $\sim$55 $R_\odot$ (and $T(r)$ remains 
$\gtrsim 10^4$ K out to $\gtrsim 20~R_\odot$).  These radii are larger than the
size of the donor star while it is near the main sequence or subgiant branch
(i.e., while it is hot).  Thus, it is quite possible that the largest contribution to 
the optical light from the IMBH systems comes from the intrinsic radiation 
emitted by the accretion disk itself.  We will see the effect of this in \S3 
(see especially Fig.\,\ref{fig:imbh}).

\vspace{0.3cm}

\subsection{Irradiation of the Donor Star}

We do not consider the effects of X-ray irradiation of the donor star.  For 
our IMBH models the half angle subtended by the donor star ranges 
between $\sim$5$^\circ$ and 10$^\circ$.  Since we have assumed an accretion
disk with only a modest half angle of 6$^\circ$, most or all of the donor star
can be expected to be shielded by the accretion disk.  For the LMBH models
the donor stars, at least initially, subtend larger half angles of $\sim$18$^\circ$ to 24$^\circ$.  
The fractional solid angle that the donors subtend ranges between $\sim$0.025
and 0.036 while they are on the main sequence (and even less when they are
on the giant branch and have transferred a significant amount of their mass).
After taking into account shadowing by the accretion disk, the albedo of the
donor star surface, and the Lambertian angular dependence of the disk
irradiator, we find that no more than 0.0018 $\rightarrow$ 0.0026 of the 
X-ray luminosity is reprocessed on the face of the donor star.  This is less
than comes from the irradiated accretion disk and, furthermore, the larger distance
between the X-ray source and the donor compared to that of the irradiated
disk renders the flux even smaller yet.  Therefore, for this work, we have
neglected the irradiation of the donor stars.

\begin{figure*}[h]
\centering
\includegraphics{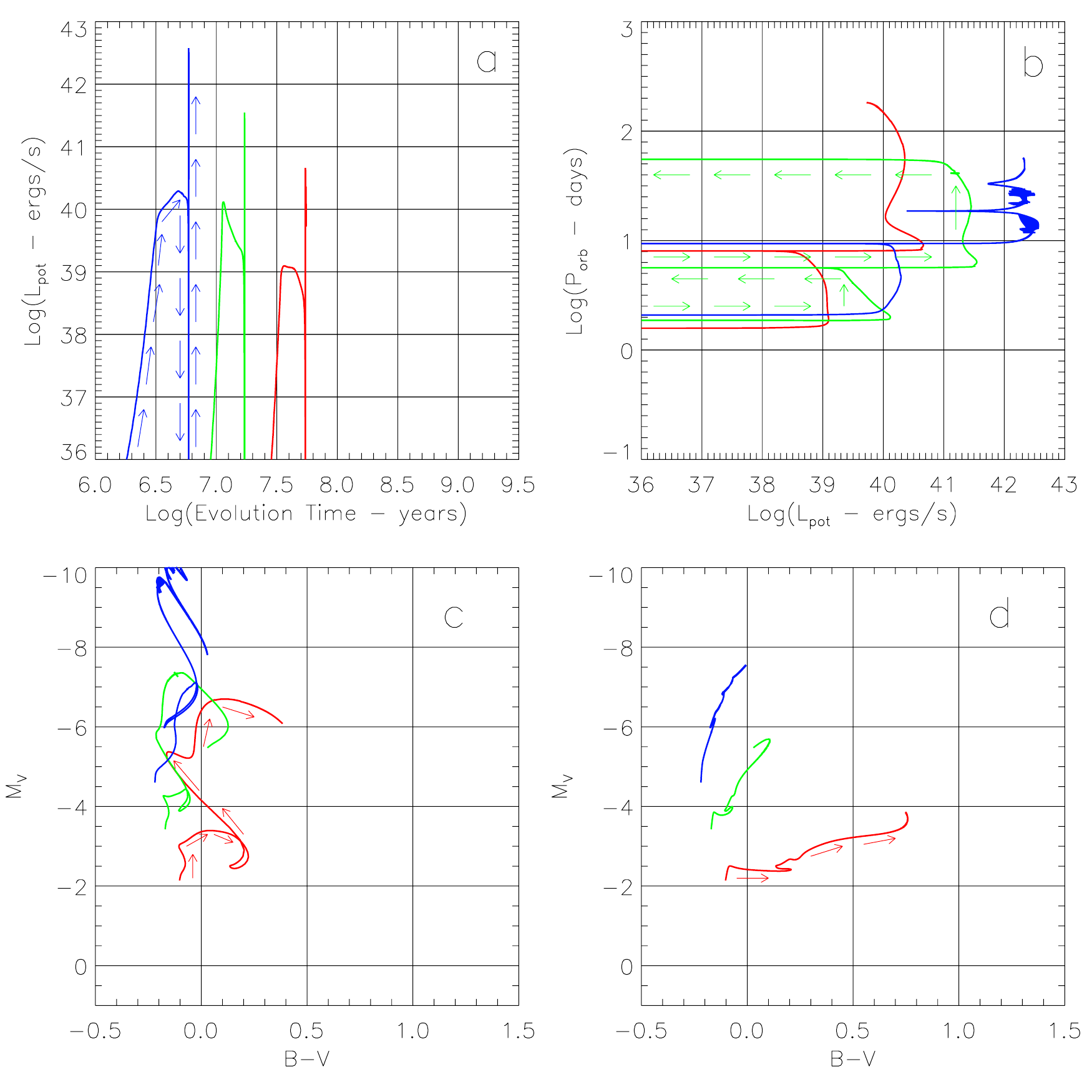}
\caption{Evolutionary tracks of illustrative binary evolution calculations. Panel 
(a): evolution of X-ray luminosity with age of the system.  Panel 
(b): Orbital period versus X-ray luminosity.  Panel (c): 
Color-magnitude diagram for the binary system. Panel (d): 
Color-magnitude diagram of the donor star alone. The blue curves show 
the evolution of an IMBH binary, typical of Model Ic, with an initial 
donor mass ($M_{\rm don}$) of 30 $M_\odot$, a black hole mass 
($M_{\rm BH}$) of 1000 $M_\odot$, and an initial orbital period 
($P_{\rm orb}$), of $2.1$ days. The green curves represent Model Lb 
with $M_{\rm don} = 13\,M_\odot$,  $M_{\rm BH} = 12\,M_\odot$, 
and $P_{\rm orb}$ = $1.9$ days. And, the red curves represent Model 
La with $M_{\rm don} = 7\,M_\odot$,  $M_{\rm BH} = 
11\,M_\odot$, and $P_{\rm orb}$ = $1.6$ days.}
\label{fig:singleplots}
\end{figure*}

\begin{figure*}[t]
\centering
\includegraphics{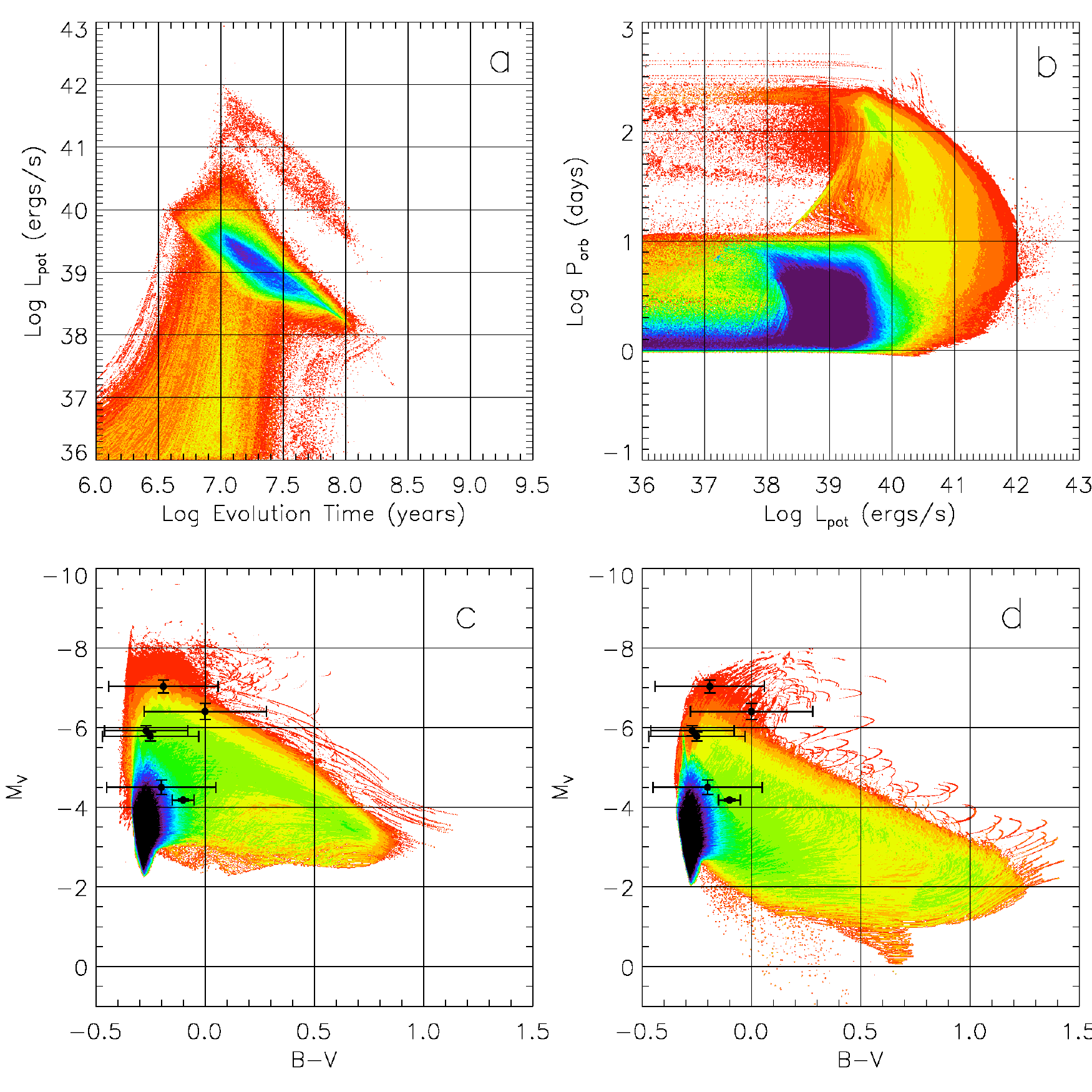}
\caption{Population diagrams for Model La, LMBH with $\lambda = 0.1$
and $q_{\rm crit}$ for the primordial binaries taking a value of 1.2 or 2.0 
(see text, eqs.~(1) and (2), and Table 2 for details). 
Panel (a): Evolution of potential X-ray luminosity with age of the 
system.  Panel (b): Orbital period versus potential X-ray luminosity. 
Panel (c): Color-magnitude diagram of the binary systems.  Panel (d): 
Color-magnitude diagram for donor stars alone. For both panels (c) 
and (d), contributions to the CMD image are made {\em only during times when} 
$L_{\rm pot} \gtrsim 2 \times 10^{39}$ ergs s$^{-1}$. Each diagram is an 
image of $700 \times 700$ pixels, and represents the 30,000 X-ray 
binary evolution calculations that we computed. For each diagram, the parameter 
values from all the tracks were registered in each pixel that was 
traversed. The intensity in panel (a) represents the number of 
systems in the pixel, whereas the intensity in each of the other 
panels represents the accumulated evolution time spent in the pixel. 
The colors are scaled according to the 1/4 root of the 
corresponding intensities, with purple being of highest intensity and 
red being the lowest; the actual ratio of values between purple and red
is $\sim$200:1. In panel (c) the irradiation is taken to be from the full
$L_{\rm pot}$.}
\label{fig:modelLa}
\end{figure*}

\begin{figure*}[t]
\centering
\includegraphics{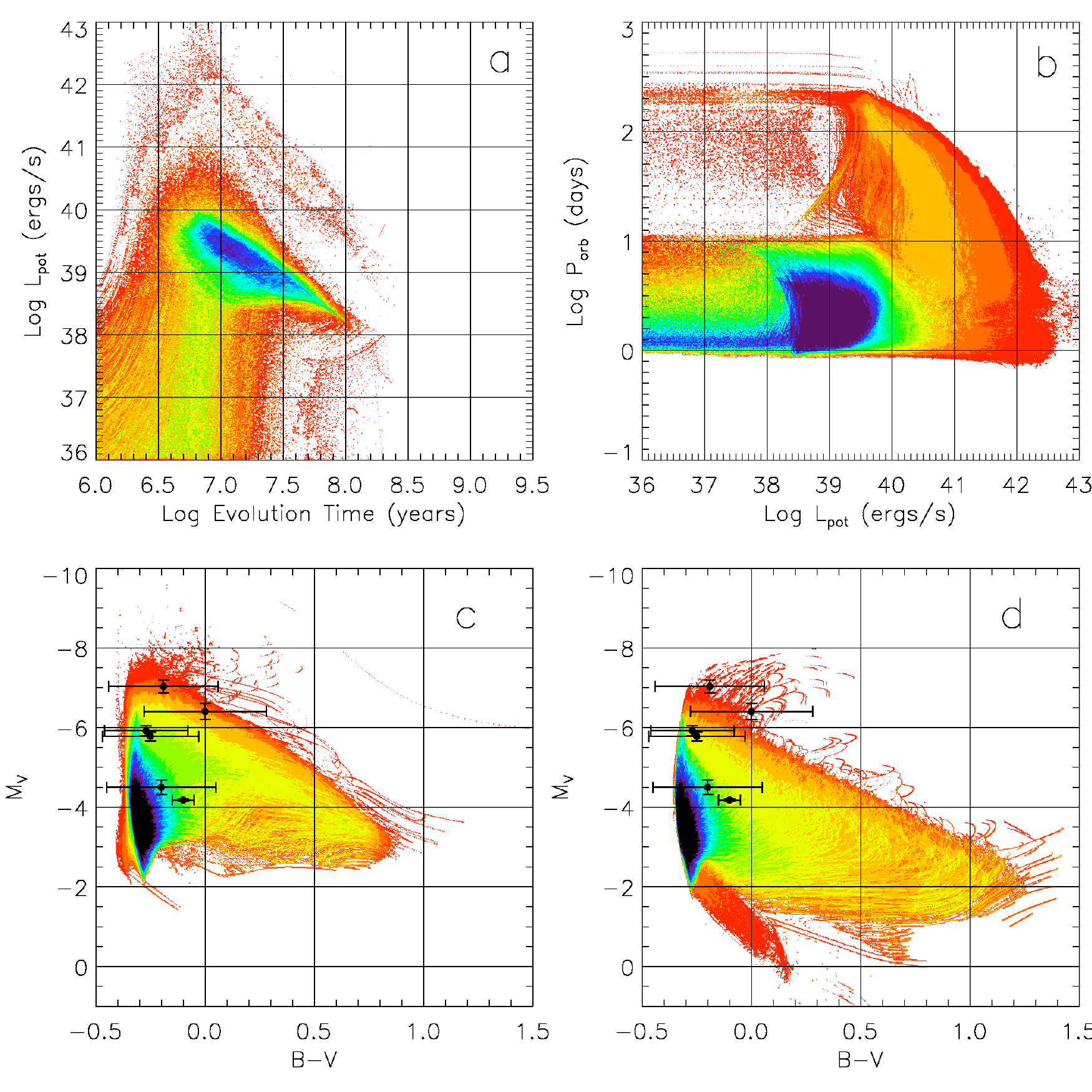}
\caption{Population diagrams for Model Lb, LMBH with $\lambda = 0.1$
and $q_{\rm crit}$ for the primordial binaries taking a value over the range
of $1.2 - 2.0$ (see text, eq.~(3), and Table 2 for details). 
Panel (a): Evolution of potential X-ray luminosity with age of the 
system.  Panel (b): Orbital period vs. potential X-ray luminosity. 
Panel (c): Color-magnitude diagram of the binary systems.  Panel (d): 
Color-magnitude diagram for donor stars alone. All other descriptors 
are the same as for Fig.\ref{fig:modelLa}.}
\label{fig:modelLb}
\end{figure*}

\begin{figure*}[t]
\centering
\includegraphics{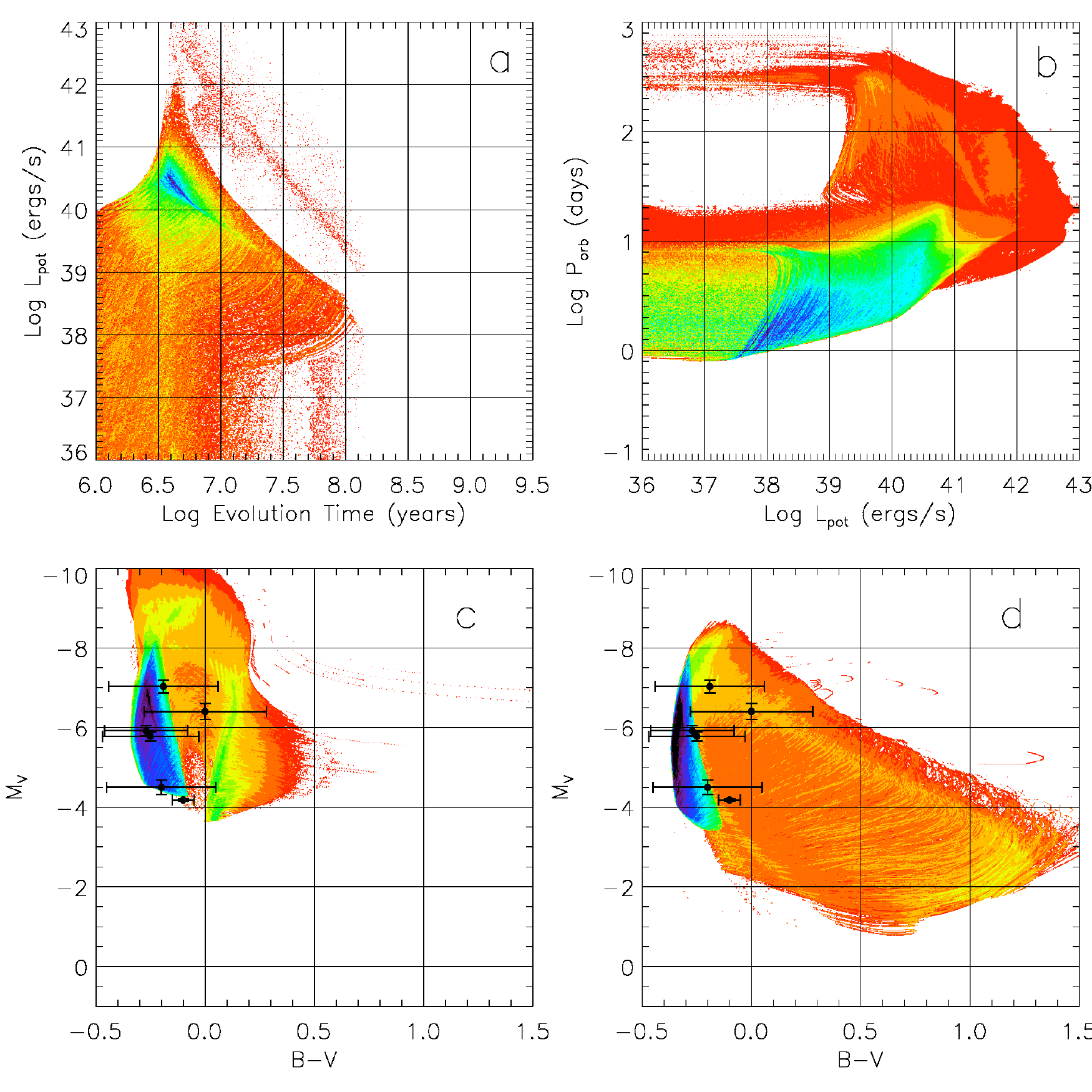}
\caption{Population diagrams for IMBH Model Ic: Panel (a): Evolution of 
potential X-ray luminosity with age of the system.  Panel (b): 
Orbital period versus potential X-ray luminosity.  Panel (c): 
Color-magnitude diagram of the binary systems.  Panel (d): 
Color-magnitude diagram for donor stars alone. All other descriptors 
are the same as for Fig.\ref{fig:modelLa}.}
\label{fig:imbh}
\end{figure*}

\vspace{0.3cm}

\section {Results and Discussion}
\label{sec:results}

We present models of ultraluminous X-ray sources including 
calculations of the optical flux from the system. The optical flux 
includes contributions from the donor star as well those due to 
radiation from the accretion disk. As mentioned previously, we have 
chosen three representative populations of binary systems for our 
study. Two of these populations have LMBHs and one population has 
IMBHs as accretors. At the beginning of each evolution, the donor 
star is on the zero age main sequence (ZAMS), and the initial system 
parameters are set by the binary population synthesis algorithms 
described above.

\subsection{Evolution of a Single Binary}

Figure \ref{fig:singleplots} shows sample evolutionary tracks 
for three individual ULX binaries. The three binaries are 
illustrative of the three models listed in Table 
\ref{tab:modelsdata}. Starting from the upper left 
panel, panel (a) shows the evolution of $L_{\rm pot}$ with the age of 
the system ($t_{\rm ev}$). Panel (b) presents the evolution of $L_{\rm 
pot}$ with the orbital period, $P_{\rm orb}$. Panel (c) shows the track in 
the color-magnitude diagram (CMD) of the ULX binary, taking into account 
the optical flux from the donor star as well as the flux due to 
radiation from the disk. And, panel (d) represents the CMD of the donor 
star alone. In calculating the optical flux from the disk, $L_{\rm 
pot}$ was used to irradiate the disk, which allows for possible 
violation of the Eddington limit. The evolution of the various 
parameters with time is denoted by the arrows. The blue curves 
represent the evolution of an IMBH binary with an initial donor mass, 
$M_{\rm don}$ = 30 $M_\odot$, a black-hole mass, $M_{\rm BH}$ = 
1000 $M_\odot$, and an initial orbital period, $P_{\rm orb}$ = $2.1$ 
days; taken from Model Ic (see Table \ref{tab:modelsdata}). The green 
curves represent Model Lb with $M_{\rm don} = 13\,M_\odot$, 
$M_{\rm BH} = 12\,M_\odot$, and $P_{\rm orb}$ = $1.9$ days. And, 
the red curves are for Model La with $M_{\rm don} = 7\,M_\odot$, 
$M_{\rm BH} = 11\,M_\odot$, and $P_{\rm orb}$ = $1.6$ days. In 
order to illustrate the direction of evolution of the tracks, on each 
panel arrows are marked on the track that is most clearly separated 
from the others.

Let us consider the $L_{\rm pot}-t_{\rm ev}$ tracks in panel (a) for 
illustration, and follow the blue curve. The evolution starts with 
the donor on the ZAMS. As the donor evolves through the main sequence 
and somewhat beyond, the radius increases slightly and the donor 
fills its Roche lobe. Mass transfer then takes place onto the accretor 
through the inner Lagrange point.
The $L_{\rm pot}-t_{\rm ev}$ curve depicts a modest increase in 
$L_{\rm pot}$ as the donor evolves through the main sequence. For 
high-mass stars, as are considered here, there is an overall 
contraction phase before the star ascends the giant branch. During 
the contraction phase, even a small decrease in radius leads to a 
large reduction in the mass-transfer rate, $\dot{M}$, and hence the 
observed dip in $L_{\rm pot}$. After the contraction phase, the star 
ascends the giant branch expanding through the Hertzsprung gap on a 
thermal timescale. This increase in radius leads to a spike in 
$L_{\rm pot}$. The effects of the various phases of evolution of the 
donor star also manifest themselves through the other properties of 
the system. For example, the development of $P_{\rm orb}$ through the 
various evolutionary phases is evident from panel (b). One can see 
the sharp increase in the orbital period as the star expands on the 
giant branch. The relatively smaller variations in $L_{\rm pot}$ as 
the period changes indicate the near constant, even if rapid, rate of 
growth of the donor star. Panels (c) and (d) depict the evolution of 
the optical properties of the system. As is apparent from panel (d), 
the increase in the absolute magnitude of the donor star as it 
ascends the giant branch takes place along with a corresponding 
decrease in its effective temperature. In the context of a binary 
system, however, some of the X-ray flux is reprocessed in the accretion disk 
and the tracks reflect the effects of mass transfer, as seen from 
panel (c).

\subsection{Population Diagrams}

Figures \ref{fig:modelLa}, \ref{fig:modelLb} and \ref{fig:imbh} show 
color images of 30,000 evolutionary tracks each, corresponding to 
Models La, Lb and Ic, respectively (see Table \ref{tab:modelsdata}). 
The panel arrangement and parameters shown are the same for all three 
figures, and match those of Fig.\,\ref{fig:singleplots}.  Each panel 
contains a 700 $\times$ 700 image matrix of the corresponding 
parameters. For panel (a), the matrix covers 7 decades in $L_{\rm 
pot}$ and 3.5 decades in evolution time, in equally spaced 
logarithmic intervals. For each step of an evolutionary track, the 
position of the $L_{\rm pot}-t_{\rm ev}$ pair is located in the 
matrix and a value of ``1'' is added to the corresponding matrix 
element. The resulting matrix, after recording all the evolution 
steps from all the 30,000 tracks, is displayed as a color image, with 
1/4-root scaling in intensity to enhance the dynamic range. For 
panel (b), the matrix covers 4 decades in $P_{\rm orb}$ and 7 decades 
in $L_{\rm pot}$, in equally spaced logarithmic intervals. Each time 
the step of an evolution track lands in a matrix element, the 
evolution time-step is added to the corresponding matrix element, as 
opposed to adding a value of `1' as was the case for panel (a). For 
panels (c) and (d), the matrices cover 11 units in the absolute 
visual magnitude, $M_V$, and 2 units in color index, in equally 
spaced linear intervals. Similar to the $P_{\rm orb}$ versus $L_{\rm 
pot}$ image in panel (b), a value equal to the evolution time step is 
added to the CMD matrix each time the step of an evolution track 
lands in a particular matrix element.  However, it is
extremely important to note that contributions to the CMD are made
{\em only during times when} $L_x \gtrsim 2 \times 10^{39}$ ergs s$^{-1}$
i.e., when the source would be a potential ULX.

As in panel (a), the matrices 
in the remaining three panels are also displayed as color images with 
1/4-root scaling in intensity (i.e., relative probability) to 
enhance the dynamic range. The colors reflect this scaling, violet 
being of highest intensity and red being the lowest.

Thus, the intensity in panel (a) is a measure of the number of 
systems with a particular $L_{\rm pot}$ found at a particular time in 
the evolution of the cluster. These numbers allow one to calculate 
the number of active ULX systems in the population at any given epoch 
and to make estimates of the numbers of ULXs of a certain $L_{\rm 
pot}$ in typical galaxies. For all the other images, the intensity at 
a point is a measure of the total duration (time) spent by all the 
systems in that interval of parameter space. Since this intensity 
incorporates both the numbers of systems and the amount of time each 
system spends in that interval of parameter space, it provides a 
probability map of the corresponding parameter space.  For purposes 
of discussion later, we shall refer to violet and dark blue regions 
in these diagrams as regions of high probability and to regions in 
red as those of low probability.

\subsection{Evolutionary Phases of the Donor Stars}

Comparing panel (c) from Fig.\,\ref{fig:singleplots}, to that of 
Figs.\,\ref{fig:modelLa}, \ref{fig:modelLb}, and \ref{fig:imbh}, we 
see that the regions of high probability 
correspond to the early phases of the donor star's evolution, i.e., on the main 
sequence or the sub-giant branch phase. This is apparent because the 
donor star spends a predominant amount of its lifetime on the main 
sequence, and evolves rather quickly on the giant branch. 
Consequently, the regions of low probability on the diagram are the 
regions corresponding to the giant-branch phase of the evolutionary 
tracks or other regions not readily accessed during the binary evolution. 
However, this description of the phases of evolution with respect to 
regions on the CMD is only approximate. The exact structure of the 
CMD is a result of a complex interplay between the evolution of the 
donor star, its mass loss, and the radiation from the disk at a 
particular evolutionary stage. 
From panel (c) of Figs.\,\ref{fig:modelLa}, \ref{fig:modelLb}, and \ref{fig:imbh}, 
we find that, with a high probability, the donor stars belong to the equivalent 
luminosity classes for single stars of IV and V, with a $B-V$ color range of 
about $-0.35$ to $-0.10$ (corresponding to an effective spectral class of 
O through late B).

In many of the panels in Figs.\,\ref{fig:modelLa}, \ref{fig:modelLb}, and 
\ref{fig:imbh} one can discern some of the individual tracks, 
roughly parallel to each other in the initial stages of the 
evolution. Tracks higher on the diagram (i.e., lower $M_{V}$) 
correspond to higher-mass donors since, at any evolution phase, more 
massive stars are brighter. Higher-mass donor stars become progressively 
rarer and also spend lesser amounts of time on the main sequence; hence 
the probability fades toward lower $M_V$.

\subsection{Relevance to the Observations}

The data from the optical observations of six ULXs discussed 
earlier (see Table 1) are plotted along with the model CMDs. For models La 
(Fig.\,\ref{fig:modelLa}) and Lb (Fig.\,\ref{fig:modelLb}), we see that two of the 
six data points fall on high-probability regions of the CMD (panel c), two on regions 
of moderate probability, and two on low-probability regions.
However, for the IMBH scenario all six data points fall on or close to the 
high-probability regions of the CMD in Fig.\,\ref{fig:imbh}. 
Thus, in the framework of the models presented here, the IMBH scenario 
appears to be the more favorable one.

We also see from Fig.\,\ref{fig:imbh} that all the observational 
points fall in regions corresponding to the very high donor masses. 
This indicates how the IMBH models fit the data better than the LMBH 
models. The brighter magnitudes of the observed systems require 
massive donor stars. And, massive donor stars need more massive 
accretors in order for mass transfer to occur stably via Roche-lobe 
overflow. In the LMBH scenario, the donor masses are limited by the fact 
that they cannot be more than about twice the mass of the BH 
accretor in order for stable mass transfer to take place.  In practice,
this limits most of the donor masses to be $\lesssim 20~M_\odot$.
In the IMBH case, on the other hand, the BH mass has been set 
at $1000\,M_\odot$ and the maximum donor mass we consider is 
$50~M_\odot$. Quite 
clearly, it is the higher-mass donors ($M \gtrsim 25\,M_\odot$) that 
lead to the high probability region where the observed systems lie. 
Such high-mass donors in the LMBH models are relatively scarce.

\subsection{Effects of Radiation from the Accretion Disk}

Panel (d) of each figure shows the CMD of the donor stars alone. 
Comparing panels (c) and (d) in each figure, we see that the 
difference is most apparent in the IMBH scenario. Comparing panels 
(c) and (d) in Fig.\,\ref{fig:modelLa}, we see that the CMDs differ 
significantly only in regions corresponding to the giant-branch 
phases of the evolutionary tracks, and remain quite similar over the
rest of the diagrams. This observation is similar for
Fig.\,\ref{fig:modelLb}.  And since the giant-branch phases lie in the
low probability regions of the diagram, one can say that the optical
flux from disk radiation contributes relatively little to the
observed optical appearances of ULXs in the LMBH scenario. By
contrast, for the IMBH model, a comparison of panels (c) and (d) in
Fig.\,\ref{fig:imbh} shows that the contribution of disk irradiation 
and {\em intrinsic} disk radiation from viscous losses (see \S 2.5)
is quite significant in this case, and has substantially altered the CMDs in 
most phases of the evolution.

The greater influences of the disk radiation for the IMBH
model (see Fig.\,\ref{fig:imbh}) compared to those for the LMBH models
seen in Figs.\,\ref{fig:modelLa} and \ref{fig:modelLb} can be explained 
by two effects.  First, in general, the IMBH systems have higher luminosities
by about an order of magnitude -- largely by virtue of their higher mass 
donor stars.  This obviously increases the power going into disk 
irradiation.  However, a comparison of IMBH and LMBH systems at 
the {\em same} values of $P_{\rm orb}$ and $L_{\rm pot}$ reveals an 
important difference.  The disks in IMBH systems still emit much more optical 
power than do the LMBH models, and this is predominantly at the
blue end of the spectrum.  The reason is that there is a substantial 
amount of intrinsic radiation from the disk (viscous release of 
gravitational potential energy; see \S2.5).  The physical explanation 
for this is straightforward.  At the same radial distance from the 
central black hole, the IMBH system releases $\sim$100 times more
gravitational potential energy than the corresponding LMBH system
(simply due to the higher mass of the black hole).  This release can end
up dominating over the disk irradiation in regions where the bulk of the 
optical emission is released (see Fig.\,\ref{fig:temp}).

Finally, it should be noted that the CMD for the donor star in a binary 
is expected to be different from that of a single star because 
of the mass loss via Roche-lobe overflow.

\subsection{Orbital Period versus X-Ray Luminosity}

Panel (b) in each of Figures \ref{fig:singleplots}, \ref{fig:modelLa}, 
\ref{fig:modelLb}, and \ref{fig:imbh} shows the evolution of $L_{\rm 
pot}$ with $P_{\rm orb}$. As can be seen in 
Fig.\,\ref{fig:singleplots}, the variations in $P_{\rm orb}$ reflect 
the changes in radius of the donor star through the various phases of 
evolution. Owing to the dependence of the Roche-lobe radius on the 
orbital separation, any rapid increase in the radius of the star 
while in contact with its Roche lobe causes a rapid increase in the 
orbital separation, and hence in $P_{\rm orb}$. The final rapid rise 
in $P_{\rm orb}$ at the highest $L_{\rm pot}$ characterizes the 
expansion of the donor on the giant branch, leading to long orbital 
periods and high mass-transfer rates. As is apparent, the long values 
of $P_{\rm orb}$ in the different models indicate the giant-branch 
phases of the donors, and the short period regions pertain to the 
main-sequence and sub-giant phases of evolution. The intensity at a 
given location in panel (b) is proportional to the number of systems 
passing through the region and the amount of time spent by the systems 
in that region. Consequently, the intensity in that region is a measure of 
the probability of finding a system at that location in parameter space. We 
then see that, for any model, the most probable regions for $L_{\rm pot} 
\gtrsim 10^{39}$ ergs s$^{-1}$ lie within a period range of $\sim$1-10 days.

It follows that the total probability of a system to be in the giant-branch 
phase is proportional to the sum of all the evolution times 
recorded in the long-period region, taken to be $P_{\rm 
orb} \gtrsim 10$ days. Similarly, the total probability of being in 
the pre-giant phase is proportional to the sum of all the evolution 
times recorded in the short-period region, i.e., $P_{\rm orb} 
\lesssim 10$ days. Thus, we can calculate the relative probability for 
a ULX donor to be in the giant-branch phase versus that to be on the 
pre-giant branch by calculating the ratios of the two sums for all 
$L_{\rm pot} \gtrsim 2 \times 10^{39}$ ergs s$^{-1}$. 
The relative probabilities thus calculated are 0.08, 0.05 and 0.15 
for Models La, Lb and Ic respectively. This reflects, in a 
quantitative way, the known fact that the systems are more likely 
to be found in the pre-giant phase than in the giant phase because of the 
relatively small amount of time they spend on the giant branch.

The implication of the $P_{\rm orb}$-$L_{\rm pot}$ population diagram 
lies in the fact that given an observed value of $L_{\rm pot}$, one 
can estimate the most probable orbital period corresponding to each 
of the three ULX scenarios presented in Figs.\,\ref{fig:modelLa}, 
\ref{fig:modelLb} and \ref{fig:imbh}. For instance, given an observed 
$L_{\rm pot} = 10^{40}$ ergs s$^{-1}$, we see from 
Figs.\,\ref{fig:modelLa}, \ref{fig:modelLb} and \ref{fig:imbh} that the 
most probable periods lie between 1--5 days, 1--5 days, and 2--10
days for the Models La, Lb and Ic respectively. Thus, the diagrams serve 
as rough guidelines showing what orbital periods are to be expected. 

\begin{figure}[t]
\begin{center}
\includegraphics[width=0.48\textwidth]{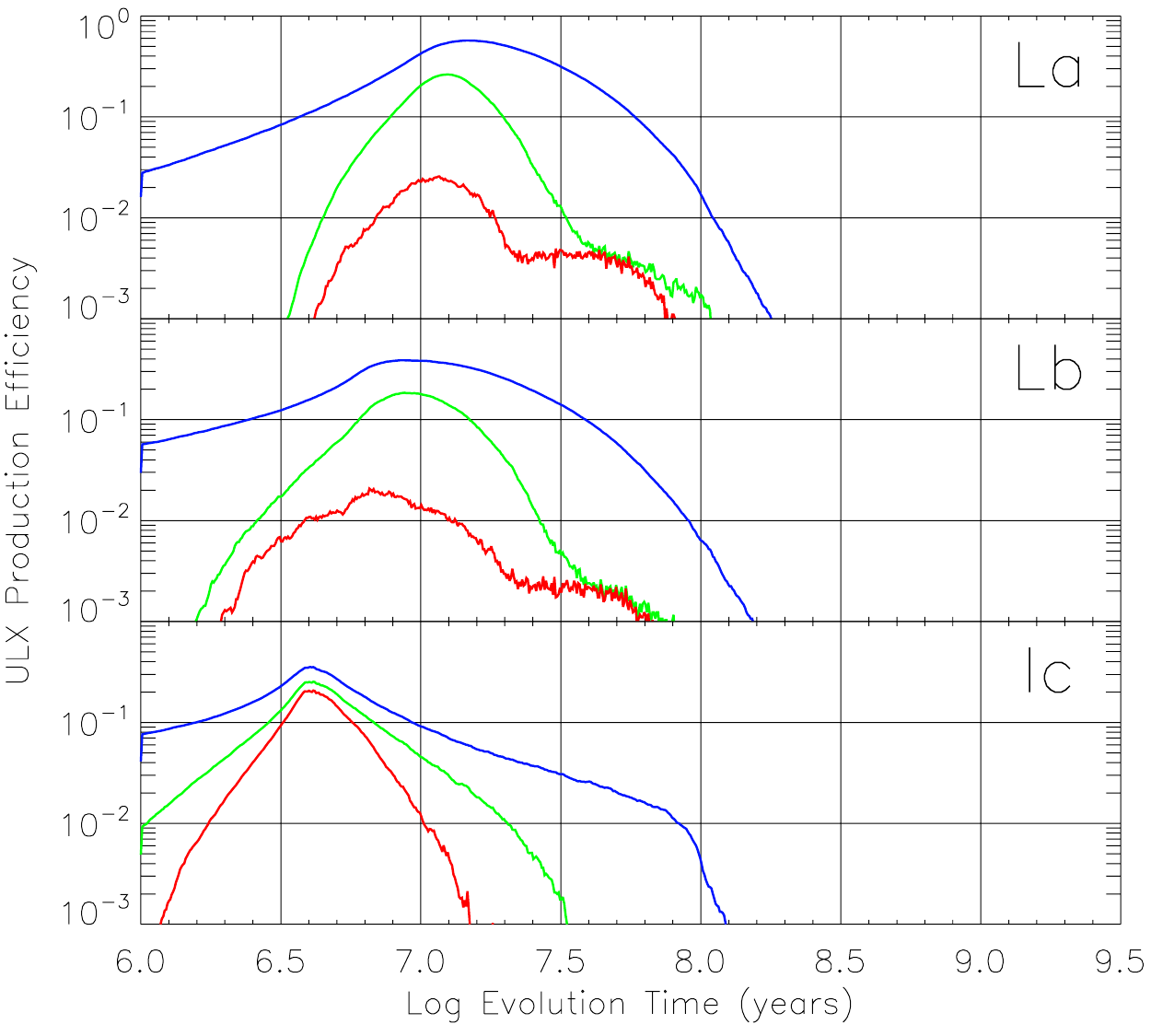}
\caption{The fraction of LMBH/IMBH binaries as a function of star 
cluster evolution time for three different lower limits on the X-ray 
luminosity. The different models are labeled in the panels. For each 
model, the blue curves represent fractions of the initial systems 
that have $L_{\rm pot} \gtrsim 10^{36}$ ergs s$^{-1}$, the green 
curves represent fractions with $L_{\rm pot} \gtrsim 2 \times 
10^{39}$ ergs s$^{-1}$, i.e., ULX luminosities, and the red curves 
are for fractions of systems with $L_{\rm pot} \gtrsim 10^{40}$ ergs 
s$^{-1}$.}
\label{fig:efficiency}
\end{center}
\end{figure}

\subsection{Evolution of the X-Ray Luminosity}

Panel (a) of each of Figs.\,\ref{fig:modelLa}, \ref{fig:modelLb} and 
\ref{fig:imbh} shows a population diagram of the evolution of the 
X-ray luminosity ($L_{\rm pot}$) with the age of the system. Defining 
the ULX luminosities as $L_{\rm pot} \gtrsim 2 \times 10^{39}$ ergs 
s$^{-1}$, we see that the systems in Model Ic are most active as ULX 
systems between $\sim$2 and 15 Myr while systems in Models La and 
Lb are most active between $\sim$4 and 25 Myr. These modest 
differences in active cluster lifetimes can be explained on 
the basis of the ranges of donor masses in the different models.
Donors with higher masses have shorter nuclear evolution timescales, 
and hence shorter overall lifetimes. So, the higher the donor masses, 
the shorter will be the active lifetimes of the systems, leading to 
shorter overall timescales over which a star cluster containing such 
binaries is X-ray active. Shorter lifetimes of systems also mean that 
the mass-transfer rates are higher, leading to systems with higher 
luminosities. Consequently, while systems in Model Ic can produce 
high $L_{\rm pot}$ systems even before ascending the giant branch, 
systems with lower masses have to wait longer to produce their peak 
X-ray luminosities. It also follows that the numbers of systems 
having high $L_{\rm pot}$ will contain a higher proportion of higher-mass 
donors. 

Figure \ref{fig:efficiency} elucidates this latter point.  It shows 
plots of fractions of the initial systems that have $L_{\rm pot}$ 
greater than $10^{36}$, $2 \times 10^{39}$ and $10^{40}$ ergs 
s$^{-1}$, at any given evolution time. We see that the maximum 
percentage of systems having (potential) ULX luminosities at any time 
during the age of a star cluster is $\sim$ 20\% for Model La, $\sim$ 30\% for 
Model Lb, and $\sim$ 25\% for Model Ic. We also see that the 
percentages of systems having $L_{\rm pot} > 10^{40}$ ergs s$^{-1}$ 
are $\sim$ 2\%, $\sim$ 3\%, and $\sim$ 20\%, respectively, for these
same models. It follows that the IMBH model allows for the largest numbers 
of very high luminosity systems. The percentages quoted above refer only to 
the most active phase of the corresponding clusters. The detailed 
evolution of the percentages with the star cluster age can be seen 
from the figure. However, the very small numbers of 
systems with $L_{\rm pot} > 10^{40}$ ergs s$^{-1}$ at {\em any} time during 
the entire cluster lifetime make the LMBH models an unlikely scenario for 
the most luminous ULXs (independent of Eddington-limit arguments).

\subsection{Estimates of ULX Numbers}

Based on the $L_{\rm pot} - t_{\rm ev}$ population diagrams, we can 
also make rough estimates of the numbers of ULXs of different $L_{\rm 
pot}$ that can be found in a galaxy in \textit{steady state}, 
assuming the different models of ULX formation.

Each element of the $L_{\rm pot} - t_{\rm ev}$ matrix contains the 
number of BH binary systems that have luminosities between $L_{\rm 
pot}$ and $L_{\rm pot} + \Delta L_{\rm pot}$ and ages between $t_{\rm 
ev}$ and $t_{\rm ev} + \Delta t_{\rm ev}$. Both, $L_{\rm pot}$ and 
$t_{\rm ev}$ are logarithmically spaced with 100 bins per decade. 
Given the fact that we evolve 30,000 binary systems for each model, 
the fraction of all high-mass binary systems that lie in the above mentioned 
intervals of $L_{\rm pot}$ and $t_{\rm ev}$ is given by $m(L_{\rm 
pot}, t_{\rm ev})/30,000$, where $m$ denotes the $700 \times 700$ 
$L_{\rm pot} - t_{\rm ev}$ matrix.

\begin{figure}[t]
\begin{center}
\includegraphics[width=0.48\textwidth]{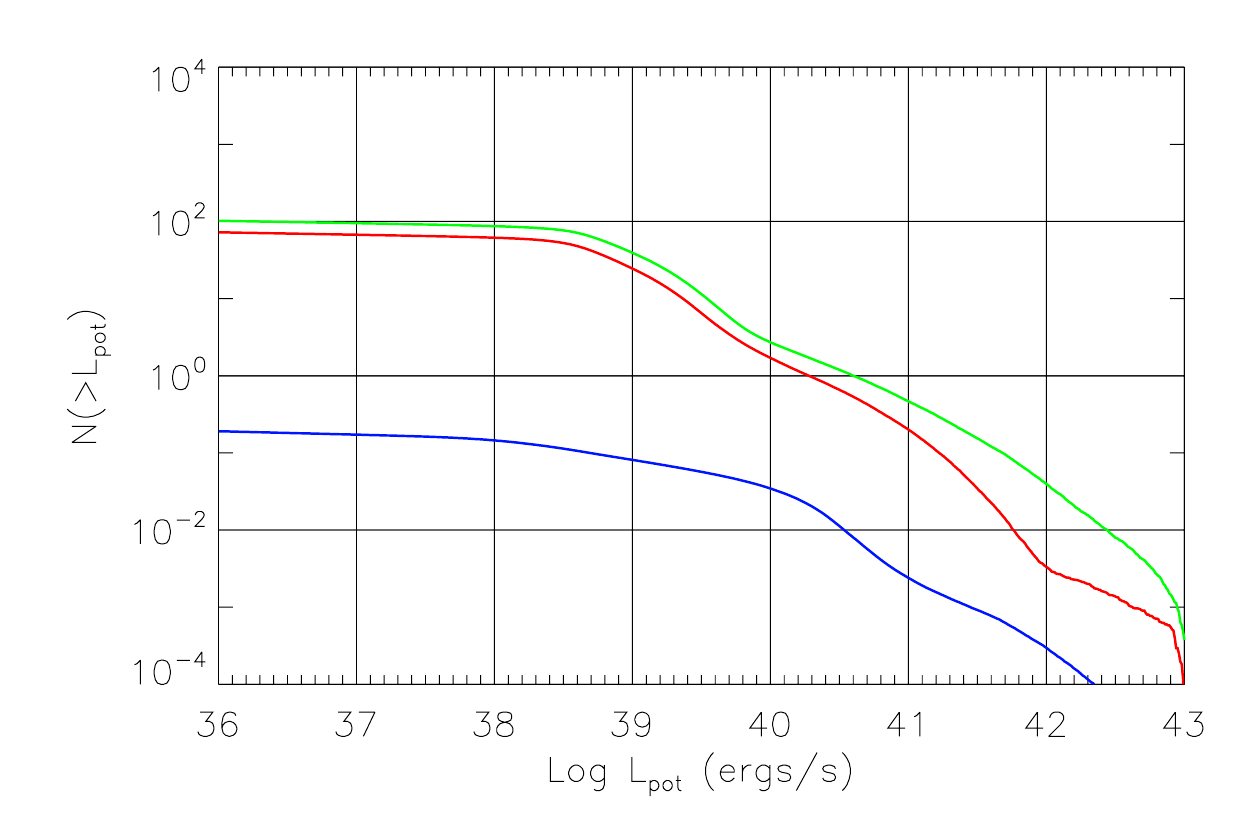}
\caption{The expected numbers of ULXs at the present epoch as a 
function of X-ray luminosity in a starburst galaxy with a core 
collapse supernova rate of 0.01 per year. The red, green, and blue 
curves correspond to Models La, Lb and Ic, respectively.}
\label{fig:numbers}
\end{center}
\end{figure}

For the LMBH models, the BH accretors in this age interval, $\Delta 
t_{\rm ev}$, have all resulted from core-collapse supernovae (SNe) in 
an equal time interval at some epoch, $t_{\rm ev}$, in the past. 
Assuming a uniform SN core-collapse rate of $R_{\rm SN}$, the total 
number of BH binaries formed in this time interval is given by 
$R_{\rm SN} \times f_{\rm BH} \times \Delta t_{\rm ev}$, where 
$f_{\rm BH}$ is the fraction of all core-collapse supernovae that 
result in BH binaries with high mass donors (as determined from the 
BPS code). It then follows that the number of binary systems with 
luminosities between $L_{\rm pot}$ and $L_{\rm pot} + \Delta L_{\rm 
pot}$ and evolutionary ages between $t_{\rm ev}$ and $t_{\rm ev} + 
\Delta t_{\rm ev}$ is given by:
\begin{equation}
\Delta N(L_{\rm pot,j},t_{\rm ev,i}) = \frac{R_{\rm SN} \times f_{\rm 
BH}}{30,000} \times \Delta t_{\rm ev,i} \times m(L_{\rm pot,j},t_{\rm 
ev,i})~,
\label{rateLmbh}
\end{equation}
where $j$ and $i$ label the matrix element bins of $L_{\rm pot}$ and 
$t_{\rm ev}$, respectively. 
The total number of sources with $L_x > L_{\rm pot}$ in a galaxy in 
steady-state can be obtained by summing over all the time bins of 
$t_{\rm ev}$, and all the luminosity bins greater than $L_{\rm pot}$, 
as:
\begin{equation}
\begin{split}
N(>L_{\rm pot,k}) = & \frac{R_{\rm SN} \times f_{\rm BH}}{30,000} \times \\
& \sum_{j=k}^{700} \sum_{i=1}^{700}\Delta t_{\rm ev,i} \times 
m(L_{\rm pot,j},t_{\rm ev,i}) ~.
\end{split}
\label{rate1}
\end{equation}
For illustration, we choose a uniform supernova rate $R_{\rm SN} = 
0.01$ yr$^{-1}$ (e.g., to represent a typical Milky-Way type galaxy), and $f_{\rm 
BH} \simeq 3.75 \times 10^{-4}$ and $9.44 \times 10^{-4}$ for models La and Lb, 
respectively. These values of $f_{\rm BH}$ are deduced from the results of our BPS 
code. The computed steady-state cumulative luminosity functions for Models La and Lb 
are shown in the red and green curves of Fig.\,\ref{fig:numbers}, respectively.

For the IMBH model, we consider the IMBHs to be formed dynamically in 
young star clusters. Following a formulation similar to 
eq.~(\ref{rate1}) for LMBH models, we find for the IMBH model:
\begin{equation}
\begin{split}
N(>L_{\rm pot,k}) = & \frac{R_{\rm ysc} \times f_{\rm IMBH} \times 
f_{\rm cap}}  {30,000} \times \\
& \sum_{j=k}^{700} \sum_{i=1}^{700}\Delta t_{\rm ev,i} \times 
m(L_{\rm pot,j},t_{\rm ev,i})~,
\end{split}
\label{rate2}
\end{equation}
where, $R_{\rm ysc}$ is the formation rate of young star clusters (with 
$M_{\rm clus} \gtrsim 10^4~M_\odot$) in a typical spiral galaxy, 
$f_{\rm IMBH}$ is the fraction of all such clusters that form an IMBH 
and $f_{\rm cap}$ is the fraction of all such IMBHs that capture a massive 
companion into a close orbit.  As illustrative values, we choose a uniform 
$R_{\rm ysc} = 10^{-5}$ yr$^{-1}$ (estimated from the combined work of 
de Grijs et al.~2003; Hunter et al.~2003; Grimm et al. 2003), 
$f_{\rm IMBH} \simeq 0.1$, and $f_{\rm cap} \simeq 0.05$ (see, e.g., 
Blecha et al.\,2006). The distribution of the numbers of systems as a 
function of $L_{\rm pot}$ is shown by the blue curve in Fig.\,\ref{fig:numbers}.

It can be seen from Fig.\,\ref{fig:numbers} that the expected numbers
of ULX systems per galaxy are in fair agreement with the observations
(Ptak \& Colbert 2004; Grimm et al.~2003) for the IMBH model, albeit
with very large uncertainties. Specificially, these estimates in
Fig.\,\ref{fig:numbers} indicate that a typical galaxy is expected to
harbor $\sim$0.07 and $\sim$0.03 ULXs with $L_{\rm pot} > 2 \times
10^{39}$ and $10^{40}$ ergs s$^{-1}$, respectively.  By contrast, for
the LMBH models, the expected numbers range from about $\sim$$12-20$
for $L_{\rm pot} > 2 \times 10^{39}$ ergs s$^{-1}$ per galaxy to
$\sim$$1.4-2.7$ with $L_{\rm pot} > 10^{40}$ ergs s$^{-1}$.  These
numbers seem to be too large by about a factor of $\sim$100 compared with the
observations (Ptak \& Colbert 2004; Grimm et al.\,2003).  Moreover,
there is considerably less flexibility, or uncertainty, in the LMBH
models in comparison with the IMBH models.  And, while one might
dismiss the number of predicted ULXs with $L_{\rm pot} > 10^{40}$ ergs
s$^{-1}$ as violating the Eddington limit beyond the realm of the
plausible, the predicted systems with $L_{\rm pot} >2 \times 10^{39}$
ergs s$^{-1}$ are not so easily dismissed.  One completely
straightforward way of reducing the computed numbers of ULX systems in
the LMBH model is to adjust the $\lambda$ parameter downward (recall
that $\lambda$ is the dimensionless inverse binding energy of the
stellar envelope of the BH progenitor).  From the output of our BPS
code, and an examination of Fig.~3 in Podsiadlowski et al.~2003, we
can see that a value of $\lambda$ in the range $0.01-0.03$ will
decrease the rate of production of BH binaries with high-mass donor
stars by factors of $\sim$$10-100$. Such values of $\lambda$ are quite
consistent with what stellar structure calculations tell us (see
Fig.~1 of Podsiadlowski et al\ 2003; Dewi \& Tauris 2000).

Overall, there are a number of uncertainties associated with the 
various parameter values that we have used to calculate the ULX 
number estimates from eqns.~(\ref{rate1}) and (\ref{rate2}). The 
primary utility of Fig.~\ref{fig:numbers} beyond providing crude 
estimates is that, given more accurate values for the uncertain input 
parameters, one can find the corresponding expected numbers of 
ULXs by simply scaling the expected numbers inferred from the figure.

\section{Summary and Conclusions}

In this paper, we have explored the optical properties and other 
observable parameters for a range of ULX models. We investigated a 
very large number of systems from three different ULX populations, 
two consisting of stellar-mass BH accretors, which we refer to as 
stellar-mass (i.e., ``low-mass'') black holes (LMBHs), and one 
consisting of intermediate-mass black-hole accretors (IMBHs). For 
each population, we computed the evolution of 30,000 individual 
binary systems, and generated population diagrams to explore a wide 
region of parameter space. We computed optical CMDs for the 
donor stars and for the binary systems including 
radiation from the accretion disk. We also computed population 
diagrams showing the evolution of $L_{\rm pot}$ with donor age, and 
those showing the evolution of $L_{\rm pot}$ with $P_{\rm orb}$.

When we plot the observed colors and magnitudes for six ULX systems 
found in the literature on the model CMDs, we find that all the data points 
lie in or near the high-probability regions spanned by the IMBH models 
on the CMD. On the other hand, fewer than half of the observed systems 
lie close to the high-probability region spanned by the LMBH 
models La and Lb, respectively.
In light of this observation, we conclude that IMBH models are 
somewhat favored by the optical observations.  The locations 
of the data on the CMDs indicate that these systems correspond to 
high mass donors, with $M_{\rm don} \gtrsim$ 25 $M_\odot$. This 
provides an indication as to why the IMBH models are favored over 
LMBH models. Our BPS calculations for LMBH binaries do not produce
very many successful systems with a donor mass of $\gtrsim 20~M_\odot$, 
owing to limitations on the mass 
of the BH ($\sim$$6-15~M_\odot$) and the requirement of stable 
mass transfer. On the other hand, the mass of the BH accretor in the 
IMBH models is set at 1000 $M_\odot$, allowing for very massive donor 
stars, and hence the ability to account for higher optical 
luminosities.

We also find from the CMDs that the regions of high probability 
correspond to the initial phases of evolution of the donors. This is 
apparent because all stars spend a predominant fraction of their 
lifetimes on the main sequence. However, this is particularly 
consistent with the color-magnitude observations, all of which lie in or 
very near the high probability regions of the IMBH models. This leads us 
to conclude that the donor stars need to be predominantly on the main 
sequence or on the sub-giant branch. The $B-V$ color range of about 
$-0.35$ to $-0.10$, corresponding to the high probability regions on 
the CMDs, indicates that the effective spectral class of the donors 
should be O through late B.  We also discuss the effects of 
X-ray irradiation of the accretion disk, as well as intrinsic radiation generated 
viscously in the disk, on the different models. Disk radiation is found to 
be significant in the IMBH models, but less so in the LMBH models. We 
suggest that the main effects operating here are: (i) the higher donor masses 
in the IMBH systems, and hence higher values of $L_{\rm pot}$, and (ii)
the larger gravitational potential energy released at a given radial distance
in the accretion disk in the IMBH models.

We compute population diagrams showing the evolution of $L_{\rm pot}$ 
with $P_{\rm orb}$. As concluded previously from the CMDs, the 
$P_{\rm orb}$--$L_{\rm pot}$ diagrams show that the giant 
branch phases of evolution in these systems form a low-probability 
region in parameter space. It is seen that the most probable periods 
for all the models lie in the range of $\sim$$1-10$ days, corresponding to the 
main-sequence and sub-giant phases of evolution. We quantify the 
likelihood of ULXs being in the pre-giant phase by estimating the 
relative probabilities for the systems to be in a pre-giant phase 
versus the giant phase. We find the probabilities to be in the ratios 
0.08, 0.05 and 0.14 for Models La, Lb and Ic, respectively. We 
subsequently discuss the utility of the $P_{\rm orb}$--$L_{\rm pot}$ 
diagrams for estimating the probable range of $P_{\rm orb}$ for a system 
with a known $L_{\rm pot}$.

Finally, we compute population diagrams showing the evolution of 
$L_{\rm pot}$ with the evolution time of the host star cluster. 
Similar diagrams have been reported for various ULX models in our 
previous studies (see Rappaport et al.\,2005; Madhusudhan et 
al.\,2006). In the present paper, we use the diagrams to calculate 
the ULX production efficiency as a function of the age of a star 
cluster, for the different models. We also estimate the numbers of 
ULXs in steady state for a typical galaxy as a function of $L_{\rm 
pot}$. We find, with crude estimates, that the  
IMBH models explain the observed numbers of ULXs reasonably well, 
with $\sim$$0.07$ with $L_{\rm pot} > 10^{39}$ ergs 
s$^{-1}$ per Milky Way type galaxy to $\sim$$0.03$ per galaxy with 
$L_{\rm pot} > 10^{40}$ ergs s$^{-1}$. 
For the LMBH models, the numbers range from about $\sim$$12-20$ 
for  $L_{\rm pot} > 2 \times 10^{39}$ ergs s$^{-1}$ per galaxy 
to $\sim$$1.4-2.7$ with $L_{\rm pot} > 10^{40}$ ergs s$^{-1}$.  As discussed
in \S3.8 these numbers are too large by about a factor of 100 compared
with the observations (Ptak \& Colbert 2004; Grimm et al.\,2003).  

There are at least three ways of reducing the computed numbers of
ULX systems in the LMBH model.  One straightforward way is to adjust the 
$\lambda$ parameter downward (see eq.~[4] for a definition).  We find that a 
value of $\lambda$ in the range $0.01-0.03$ will decrease the rate of production 
of BH binaries by the correct factors.  Such small values of $\lambda$ are quite
consistent with calculations of envelope binding energies of massive stars (see Fig.~1 
of Podsiadlowski et al. 2003; Dewi \& Tauris 2000).  Another possible explanation 
for the apparent overproduction of very luminous X-ray sources in the case of 
stellar-mass black-hole accretors is that the X-ray luminosities are, in fact, 
constrained to the Eddington limit, and that the large amounts of radiation-pressure 
ejected material severely attenuate the soft X-rays coming from the vicinity of the 
black hole.  Finally, we mention the possibility that
for case B mass transfer (i.e., when the primary progenitors of the black holes
are in the Hertzsprung gap during the time when mass transfer commences) 
the primary is more likely to produce a neutron star rather than a black hole 
even for fairly high initial masses (e.g., $50-60~M_\odot$). As has first been argued by 
Brown et al.\ (1999) and then has been confirmed in detailed calculations by Brown 
et al.\ (2001),  this has to do with differences during helium core burning.  
Basically the final structure of a star is very different depending on whether
the star burns helium with a hydrogen-burning shell around it or
without the hydrogen-burning shell (as one would expect if the
star loses its envelope before or early during helium burning).
Without the hydrogen-burning shell the star ends up with a much smaller
iron core at the end and most likely results in the production of a neutron star.
This may be another reason why we do not find LMBH ULXs in nature.

In estimating the numbers of ULXs, we have allowed for violation of 
the Eddington limit by using $L_{\rm pot}$ in our calculations. This 
relaxation is imperative in order to explain ULXs using the LMBH 
models. However, the scenario by which the Eddington limit could be 
violated in a BH binary is not clear. We note that the numbers for 
the LMBH models with $L_{\rm pot} > 10^{40}$ ergs s$^{-1}$ could be 
quite unphysical because of the fact that for these luminosities 
$L_{\rm pot} \gg L_{\rm Edd}$. The Eddington limit for a $20~M_\odot$ 
accretor is $\sim$$2.5 \times 10^{39}$ ergs s$^{-1}$. Allowing for the 
fact that a massive donor star could be transferring helium onto the 
accretor during the later stages of its evolution, the Eddington 
limit is increased by a factor of 2, to $\sim$$5 \times 10^{39}$ ergs s$^{-1}$. 
Considering, however, that the bolometric luminosity may well be a factor of 
$\sim$$2-3$ times the X-ray luminosities observed in the $1-10$ kev 
X-ray band, the effective Eddington limit in the X-ray band is only $\sim$$1.5 \times 
10^{39}$ ergs s$^{-1}$. Thus, even if we allow for a violation of the Eddington limit by 
reasonable factors of a few, and thereby achieve $L_{\rm pot}$ up to  
$\sim$$10^{40}$ ergs s$^{-1}$, it is difficult to understand X-ray-band
luminosities above $\sim$$5 \times 10^{39}$ ergs s$^{-1}$. 

As for the IMBH model, we have not studied in any detail the 
formation scenarios of IMBHs or of the capture of binary companions 
by the IMBHs (but see Portegies Zwart et al.~2004; Tutukov \& Fedorova 2005; 
Blecha et al.~2006). Despite much evidence in the literature supporting IMBH 
binaries as candidates for ULXs, the formation mechanisms of IMBHs 
are not clear. In particular, it is unclear how supermassive stars evolve,
and whether they can undergo core collapse to form an IMBH.  In calculating 
the numbers of ULXs in this scenario, 
we have assumed plausible values for the formation rate of young star 
clusters in a galaxy, the fraction of these clusters that 
successfully produce IMBHs, and the probability of capture of a 
massive companion by the IMBH.

Finally, we should keep in mind the possibility that neither the IMBH nor
LMBH binary models may be correct in explaining the most luminous
ULXs.  It is conceivable that the most luminous ULXs may, in fact, be free 
floating IMBHs that are fed by tidally disrupted passing field stars (see
Volonteri \& Perna 2005, for a related scenario).

\acknowledgements
We are grateful to Bill Paxton for all his assistance with the {\em EZ} stellar evolution code.
We thank Wlodek Klu\'zniak for helpful discussions regarding accretion disks.
SR acknowledges support from NASA Chandra Grant TM5-6003X. LN 
thanks NSERC (Canada) and the CRC Program for financial support, and 
also acknowledges the CCS at the Universite de Sherbrooke for their 
technical assistance.

\end{document}